\documentclass[letterpaper, 10 pt, conference]{ieeeconf}  % Comment this line out if you need a4paper

\IEEEoverridecommandlockouts                              % This command is only needed if 
\usepackage{graphics} % for pdf, bitmapped graphics files
\usepackage{epsfig} % for postscript graphics files
\usepackage{amsmath} % assumes amsmath package installed
\usepackage{amssymb}  % assumes amsmath package installed
\usepackage{color,xcolor}
\usepackage{multirow}
\usepackage{graphicx}
\usepackage{subfig}
\usepackage{epstopdf}
\usepackage[ruled,vlined]{algorithm2e}

\usepackage{mathtools,float}
\usepackage{enumerate}

\newtheorem{definition}{Definition}
\newtheorem{theorem}{Theorem}
\newtheorem{lemma}{Lemma}
\newtheorem{remark}{Remark}
\newtheorem{assumption}{Assumption}
\newtheorem{stassumption}[assumption]{Standing Assumption}
\newtheorem{proposition}{Proposition}

\newcommand{\argmin}{\mathop{\mathrm{argmin}}\limits}

\newcommand{\Id}{ \textrm{Id}}
\newcommand{\bR} { {\mathbb R}}
\newcommand{\bN} { {\mathbb N}}

\newcommand{\fix} { {\mathrm{fix}}}
\newcommand{\zer} { {\mathrm{zer}}}
\newcommand{\diag} { {\mathrm{diag}}}

\newcommand{\ca}[1]{\mathcal{#1}}

\newcommand{\bld}[1]{\boldsymbol{#1}}
\newcommand{\1}{\bld 1}
\newcommand{\0}{\bld 0}
\newcommand{\col }{\mathrm{col}}
\newcommand{\iffs}{\Leftrightarrow}

\title{\LARGE \bf An asynchronous, forward-backward, distributed generalized Nash equilibrium seeking algorithm.
}
\author{Carlo Cenedese$^{1}$ \and Giuseppe Belgioioso$^{2}$ \and Sergio Grammatico$^{3}$ \and Ming Cao$^{1}$% <-this % stops a space
\thanks{$^{1}$ Engineering and Technology Institute Groningen (ENTEG), Faculty of Science and Engineering, University of Groningen, The Netherlands         {\tt\small c.cenedese@rug.nl} and {\tt\small m.cao@rug.nl}. The work of Cenedese, and Cao was supported in part by the European Research Council (ERC-CoG-771687) and the Netherlands Organization for Scientific Research (NWO-vidi-14134).
}%
\thanks{$^{2}$ Control System group, TU Eindhoven, 5600 MB Eindhoven, The Netherlands
{\tt\small g.belgioioso@tue.nl}.
}
\thanks{$^{3}$ Delft Center for Systems and Control, TU Delft, The Netherlands
{\tt\small s.grammatico@tudelft.nl}. The work of Grammatico was partially supported by NWO, under research projects OMEGA (TOP 613.001.702) and P2P-TALES (ESI-BIDA 647.003.003) and by the ERC under research project COSMOS (ERC-StG 802348).}
}

\begin{document}

\maketitle
\thispagestyle{empty}
\pagestyle{empty}

%%%%%%%%%%%%%%%%%%%%%%%%%%%%%%%%%%%%%%%%%%%%%%%%%%%%%%%%%%%%%%%%%%%%%%%%%%%%%%%%
\begin{abstract}
In this paper, we propose an asynchronous distributed algorithm for the computation of generalized Nash equilibria in noncooperative games, where the players interact via an undirected communication graph. Specifically, we extend the paper ``Asynchronous distributed algorithm for seeking generalized Nash equilibria" by Yi and Pavel: we redesign the asynchronous update rule using auxiliary variables over the nodes rather than over the edges. This key modification renders the algorithm scalable for highly interconnected games. The derived asynchronous algorithm is robust against delays in the communication and it eliminates the idle times between computations, hence modeling a more realistic interaction between players with different update frequencies. We address the problem from an operator-theoretic perspective and design the algorithm via a preconditioned forward-backward splitting. Finally, we numerically simulate the algorithm for the Cournot competition in networked markets.
\end{abstract}

%%%%%%%%%%%%%%%%%%%%%%%%%%%%%%%%%%%%%%%%%%%%%%%%%%%%%%%%%%%%%%%%%%%%%%%%%%%%%%%%
\section{INTRODUCTION}
\subsection{Motivation and literature overview}
Noncooperative generalized games over networks is currently a very active research field, due to the spreading of multi-agent network systems in modern society. Such type of games emerge in several application domains, such as smart grids  \cite{dorfer:simpson-porco:bullo:16,parise:colombino:grammatico:lygeros:14}, social networks \cite{grammatico:18tcns} and robotics \cite{martinez:bullo:cortes:frazzoli:07}.      
In a game setup the players, or agents, have a private and local objective function that depends on the decisions of some other players, which shall be minimized while satisfying both local and global, coupling, constraints. Typically each agent defines its decision, or strategy, based on some local information exchanged with a subset of other agents, called neighbors. One popular notion of solution for these games is a collective equilibrium where no player benefits from changing its strategy, e.g. a \textit{generalized Nash equilibrium} (GNE). Various authors proposed solutions to this problem \cite{pavel2017:distributed_primal-dual_alg,grammatico:18tcns,facchinei:fischer:piccialli:07}. These works propose only synchronous solutions for solving noncooperative games. 
So, all the agents shall wait until the slowest one in the network completes its update, before starting a new operation. This can slow down the convergence drammatically, especially in large scale and heterogeneous systems. On the other hand, adopting an asynchronous update reduces the idle times, increasing efficiency. In addition,  it can also speed up the convergence, facilitate the insertion of new agents in the network and even increse robusteness w.r.t. communication faults \cite{BERTSEKAS:1991:Survey_Asynch}. %For these reasons, we focus in this paper into an asynchronous solution for noncooperative games.  
The pioneering work of  Bertsekas and Tsitsiklis \cite{bertsekas:1989:parallel_optimization} can be considered the starting point of the literature on parallel asynchronous optimization. 
During the past years, several asynchronous algorithms for distributed convex optimization were proposed \cite{recht:2011:hogwild,Combettes:2015:stoch_quadi_fejer,liu:2015:asynchronous_parallel_stoch_coord_desc,nedic:2011:asynchronous_broadcast-based_convex_opt}, converging under different assumptions. The novel work in \cite{peng2016arock}, provides a simple framework (ARock) to develop a wide range of iterative fixed point algorithms based on nonexpansive operators and it is already adopted in \cite{Pavel:Yi:2018:Asynch} to seek variational GNE seeking under equality constraints and using edge variables.
 % ,Wu:Yuan:Sayed:2016:Decentralized_consensus_optimization
%It uses a stochastic update of the variables that provides robustness against delayed information.
%This framework was used in \cite{QinWu:2017:Accelerated_image_proc} to develop an ADMM algorithm to speed up the image processing for the electroencephalography inverse problem. More recently, it was also adopted in \cite{Pavel:Yi:2018:Asynch} to seek variational GNE seeking under equality constraints and using edge variables.
%paper contribution
%\change{ Paper contribution:\begin{itemize}
%\item use of node variables for scalability 
%\item consider the case of inequality constraints
%\item same delays in the case between variables
%\end{itemize}   }

In this paper, we propose an extension of the work in \cite{Pavel:Yi:2018:Asynch}. Specifically, we consider inequality coupling constraints and use a restricted set of auxiliary variables, namely, associated with the nodes rather than with the edges. Especially this latter upgrade  is non-trivial and presents technical challenges in the asynchronous implementation of the algorithm, which we overcome by analyzing the influence of the delayed information on the update of the auxiliary variables. The use of node variables only, rather than edge variables, preserves the scalability of the algorithm, with respect to the number of nodes. 
\subsection{Structure of the paper}
The paper is organized as follows: Section~\ref{sec:problem_formulation} formalizes the problem setup and introduces the concept of \textit{variational} v-GNE. In  Section~\ref{sec:synch_case} the iterative algorithm for v-GNE seeking is derived for the synchronous case. The asynchronous counterpart of the algorithm is presented in Section~\ref{sec:asynch_case}. Section~\ref{sec:simulations} is dedicated to the simulation results for the problem of Cournot competition in networked markets. Section~\ref{sec:conclusion} ends the paper presenting the conclusions and the outlooks of this work.  
\section{NOTATION}
\label{sec:notations}
\subsection{Basic notation}
The set of real, positive, and non-negative numbers are denoted by $\mathbb{R}$, $\mathbb{R}_{>0}$, $\mathbb{R}_{\geq 0}$, respectively; $\overline{\mathbb{R}}:=\mathbb{R}\cup \{\infty\}$. The set of natural numbers is $\mathbb{N}$. For a square matrix $A \in \bR^{n\times n}$, its transpose is $A^\top$, $[A]_{i}$ is the $i$-th row of the matrix and  $[A]_{ij}$ represents the elements in the row $i$ and column $j$. $A\succ 0 $ ($A\succeq 0 $)  stands for positive definite (semidefinite) matrix,  instead $>$ ($\geq$) describes element wise inequality. $A\otimes B$ is the Kronecker product of the matrices $A$ and $B$.   The identity matrix is denoted by~$I_n\in\bR^{n\times n}$. $\0$ ($\1$) is the vector/matrix with only $0$ ($1$) elements. For $x_1,\dots,x_N\in\mathbb{R}^n$, the collective vector is denoted  as $\boldsymbol{x}:=\mathrm{col}(x_1,\dots,x_N)=[x_1^\top,\dots ,x_N^\top ]^\top$. 
$\diag(A_1,\dots,A_N)$ describes a block-diagonal matrix with the matrices $A_1,\dots,A_N$ on the main diagonal. The null space of a matrix $A$ is $\mathrm{ker}(A)$.
%For vectors $x,y \in \bR^n$ and a positive definite matrix $ Q \in\bR^{n\times n}$, the weighted inner product and norm are denoted by $\langle x, Qy\rangle$ and $\lVert x \rVert_{Q}=\sqrt{\langle x, Qy\rangle}$, respectively; the induced matrix norm is denoted by $\lVert A\rVert_{Q}$. For $Q=I_{n}$, the standard inner product, Euclidean norm, and Frobenius norm are obtained. %A real $n$ dimensional Hilbert space obtained by endowing $\mathcal H=(\bR^n,\lVert\cdot\rVert)$  with the product $\langle x | y\rangle_{Q}$ is denoted by $\mathcal{H}_Q$.
The Cartesian product of the sets $\Omega_i$, $i=1,\dots ,N$ is $\prod^N_{i=1} \Omega_i$.

\subsection{Operator-theoretic notation}
The identity operator is by~$\Id(\cdot)$. The indicator function $\iota_\mathcal{C}:\bR^n\rightarrow[0,+\infty]$ of $\mathcal{C}\subseteq \bR^n$ is defined as $\iota_\mathcal{C}(x)=0$ if $x\in\mathcal{C}$; $+\infty$  otherwise.
The set valued mapping $N_{\ca C}:\bR^n\rightrightarrows \bR^n$ denotes the normal cone to the set $\mathcal{C}\subseteq \bR^n$, that is $N_{\ca C}(x)= \{ u\in\bR^n \,|\, \mathrm{sup}\langle \ca C-x,u \rangle\leq 0\}$ if $x \in \ca C$ and  $\varnothing$ otherwise. The graph of a set valued mapping $\ca A:\ca X\rightrightarrows \ca Y$ is $\mathrm{gra}(\ca A):= \{ (x,u)\in \ca X\times \ca Y\, |\, u\in\ca A (x)  \}$. For a function $\phi:\bR^n\rightarrow\overline{\mathbb{R}}$, define $\mathrm{dom}(\phi):=\{x\in\bR^n|f(x)<+\infty\}$ and its subdifferential set-valued mapping, $\partial \phi:\mathrm{dom}(\phi)\rightrightarrows\bR^n$, $\partial \phi(x):=\{ u\in \bR^n | \: \langle y-x|u\rangle+\phi(x)\leq \phi(y)\, , \: \forall y\in\mathrm{dom}(\phi)\}$.  The projection operator over a closed set $S\subseteq \bR^n$ is $\textrm{proj}_S(x):\bR^n\rightarrow S$ and it is defined as $\textrm{proj}_S(x):=\mathrm{argmin}_{y\in S}\lVert y - x \rVert^2$. A set valued mapping $\ca F:\bR^n\rightrightarrows \bR^n$ is $\ell$-Lipschitz continuous with $\ell>0$, if $\lVert u-v \rVert \leq \ell \lVert x-y \rVert$ for all $(x,u)\, ,\,(y,v)\in\mathrm{gra}(\ca F)$; $\ca F$ is (strictly) monotone if $\forall (x,u),(y,v)\in\mathrm{gra}(\ca F)$ $\langle u-v,x-y\rangle \geq (>)0$ holds true, and  maximally monotone if it does not exist a monotone operator with a graph that strictly contains $\mathrm{gra}(\ca F)$. Moreover, it is $\alpha$-strongly monotone if $\forall (x,u),(y,v)\in\mathrm{gra}(\ca F)$ it holds $\langle x-y, u-v\rangle \geq \alpha \lVert x-y \rVert^2$. The operator $\ca F$ is $\eta$-averaged ($\eta$-AVG) with $\eta\in(0,1)$ if $\lVert \ca F(x)-\ca F(y) \rVert^2 \leq \lVert x-y\rVert^2-\frac{1-\eta}{\eta}\lVert (\Id-\ca F)(x)-(\Id-\ca F)(y)  \rVert^2$ for all $x,y\in\bR^n$; $\ca F$ is $\beta$-cocoercive if $\beta\ca F$ is $\frac{1}{2}$-averaged, i.e. firmly nonexpansive (FNE).
The resolvent of an operator $\ca A:\bR^n\rightrightarrows \bR^n$ is $\mathrm{J}_{\ca A} :=(\Id+\ca A)^{-1}$.

\section{Problem Formulation}
\label{sec:problem_formulation}
\subsection{Mathematical formulation}
\label{subsec:math_formulation}

We consider a set of $N$ agents (players), involved in a noncooperative game subject to coupling constraints. Each player $i\in\ca N:=\{1,\dots,N\}$ has a local decision variable (strategy) $x_i$ that belongs to its private decision set $\Omega_i\subseteq \bR^{n_i}$, the vector of all the strategies played is $\bld x:= \mathrm{col}(x_1,\dots,x_{N})\in\bR^{n}$ where $n=\sum_{i\in\ca N} n_i$, and $\bld x_{-i}=\mathrm{col}(x_1,\dots,x_{i-1},x_{i+1},\dots,x_{N})$ are the decision variables of all the players other than $i$. The aim of each agent $i$ is to minimize its local cost function $f_i(x_i,\bld x_{-i}):\Omega_i\times \Omega_{-i}\rightarrow\overline \bR$, where $\bld \Omega=\prod_{i\in\ca N} \Omega_i \subseteq \bR^{n}$, that leads to a coupling between players, due to the dependency on both $x_i$ and the strategy of the other agents in the game. In this work we assume the presence of  affine constraints between the agent strategies. These shape the collective feasible decision set
 \begin{equation}
\label{eq:collective_feasible_dec_set}
\ca{\bld X} := \bld \Omega \cap \left\{\bld x\in\bR^{n} \,|\, A\bld x\leq b \right\}\, , 
\end{equation}      
where  $A\in\bR^{m\times n}$ and $b\in\bR^m$. Then, the feasible set of each agent $i\in\ca N$ reads as  
 \begin{equation*}
\label{eq:feasible_dec_set}
\ca{X}_i(\bld{x}_{-i}) := \left\{ y\in\Omega_i \,|\, A_i y -b_i  \leq   \textstyle{\sum_{j\in \ca N\setminus\{ i\} }} b_j -A_jx_j \right\}\, , 
\end{equation*}      
where $A = [A_1,\dots,A_N]$, $A_i\in\bR^{m\times n_i}$ and $\sum_{j=1}^N b_j =b$. 
We note that both the local decision set $\Omega_i$ and how the player $i$ is involved in the coupling constraints, i.e. $A_i$ and $b_i$, are private information, hence will not be accessible to other agents.
Assuming affine constraints is common in the literature on noncooperative games \cite{Paccagnan_Gentile2016:Distributed_computation_GNE,pavel2017:distributed_primal-dual_alg}. %Furthermore, in \cite[Remark~3]{grammatico:18tcns}, it is highlighted that if the constraints are separable and convex, then they can be rewritten as affine coupling constraints.\\
In the following, we introduce some other common assumptions over the aforementioned sets and cost function.\smallskip
\begin{stassumption}[Convex constraint sets]
\label{ass:convex_constr_set}
For each player $i\in\ca N$, the set $\Omega_i$ is convex, nonempty and compact. The feasible local set $\ca X_i(\bld x_{-i})$ satisfies Slater's constraint qualification. 
\hfill \QEDopen
\end{stassumption}
\smallskip
\begin{stassumption}[Convex and diff. cost functions]
\label{ass:convex_diff_function}
For all $i\in\ca N$, the cost function $f_i$ is continuous, $\beta$-Lipschitz continuous, continuously differentiable and convex in its first argument. 
\hfill \QEDopen
\end{stassumption}
\smallskip

%The game defining the dynamics of $x_i$ for each player $i$ can be described in a compact form as a constrained best response best response w.r.t. the strategies of the other agents $\bld x_{-i}$, it reads as 
In compact form, the game between players reads as
\begin{equation}\label{eq:game_formulation}
x_i\in \argmin_{y\in\bR^n} f_i(y,\bld x_{-i})\quad \textup{s.t.} \quad y\in \ca{X}_i(\bld{x}_{-i})\:.
\end{equation}   
In this paper, we are interested in the \textit{generalized Nash equilibia} (GNE) of the game in \eqref{eq:game_formulation}.
\smallskip
\begin{definition}[Generalized Nash equilibrium]
\label{def:GNE}   
A collective strategy $\bld x^*$ is a GNE if, for each player $i$, it holds 
\begin{equation}
\label{eq:GNE_best_resp}
x_i^*\in \argmin_{y\in\bR^n} f_i(y,\bld x_{-i}^*)\quad \textup{s.t.} \quad y\in \ca{X}_i(\bld{x}_{-i}^*)\:.
\end{equation} \hfill\QEDopen
\end{definition}
\subsection{Variational GNE}
\label{sec:v_GNE}
%In Definiton~\ref{def:GNE} we introduce a broad class of equilibria for the game in \eqref{eq:game_formulation}, in the following we will introduce a particular subset of equilibria, that share some desirable characteristics.
Let us introduce an interesting subset of GNE, the set of so called \textit{variational GNE} (v-GNE), or \textit{normalized equilibrium point}, of the game in \eqref{eq:game_formulation} referring to the fact that all players share a common penalty in order to meet the constraints. This is a refinement of the concept of GNE that has attracted a growing interest in recent  years - see \cite{kulkarni:shanbhag:12} and references therein.
This set can be rephrased as solutions of a variational inequality (VI), as in \cite{facchinei:fischer:piccialli:07}.

First, we define the \textit{pseudo-gradient} mapping of the game  \eqref{eq:game_formulation} as 
\begin{equation}
\label{eq:pseudo_grad}
F(\bld x)=\col\left( \{\nabla_{x_i}f_i(x_i,\bld x_{-i})\}_{i\in\ca N}\right)\,, 
\end{equation}
that gathers all the subdifferentials of the local cost functions of the agents. The following are some standard technical assumptions on $F$, see \cite{Facchinei:2011:KKT_and_GNE,BELGIOIOSO:2017:convexity_and_monotonicity_aggr_games}.
\smallskip
\begin{stassumption}
\label{ass:subgrad_lipschitz_strong_mon}
The pseudo-gradient $F$ in \eqref{eq:pseudo_grad}  is  $\ell$-Lipschitz continuous and $\alpha$-strongly monotone, for some $\ell,\alpha>0$. \hfill\QEDopen
\end{stassumption}
\smallskip
Standing Assumption~\ref{ass:convex_diff_function} implies that $F$ is a single valued mapping, hence one can define VI($F,\bld X$) as the problem:
\begin{equation}
\label{eq:VI_GNE}
\textup{find } \bld x^*\in\bld X,\:\textup{s.t.}\: \langle F(\bld x^*),\bld x-\bld x^*\rangle \geq 0\,, \quad \forall \bld x \in \bld X \:.
\end{equation}    
Next, let us define the KKT conditions associated to the game in \eqref{eq:game_formulation}. Due to the convexity assumption, if $\bld x^*$ is a solution of \eqref{eq:game_formulation}, then there exist $N$ dual variables $\lambda^*_i\in\bR^m_{\geq 0}$, $\forall i\in \ca N$, such that the following inclusion is satisfied:
\begin{equation} \label{eq:KKT_game}
\begin{split}
\bld 0 &\in \nabla_{x_i}f_i(x_i)+A_i^\top\lambda_i^*+N_{\Omega_i}(x_i^*)\, , \; \forall i\in\ca N \\
\bld 0 &\in b-A\bld x^*+ N_{\bR^m_{\geq 0}}( \lambda^*_i)\:.
\end{split} 
\end{equation}
While in general the dual variables $\{\lambda_i\}_{i\in\ca N}$ can be different, here we focus on the subclass of equilibria sharing a common dual variable, i.e., $\lambda^*=\lambda_1^* = \dots =\lambda_N^*$. 

In this case, the KKT conditions for the VI($F,\bld X$) in \eqref{eq:VI_GNE} (see \cite{facchinei:fischer:piccialli:07,facchinei:pang}) read as   
\begin{equation} \label{eq:KKT_VI}
\begin{split}
\bld 0 &\in \nabla_{x_i}f_i(x_i)+A_i^\top\lambda^*+N_{\Omega_i}(x_i^*)\, , \; \forall i\in\ca N \\
\bld 0 &\in b-A\bld x^* + N_{\bR^m_{\geq 0}}( \lambda^*) \:.
\end{split} \:.
\end{equation}
By \eqref{eq:KKT_game} and \eqref{eq:KKT_VI}, we deduce that every solution $\bld x^*$ of VI($F,\bld X$) is also a GNE of the game in \eqref{eq:game_formulation}, \cite[Th.~3.1(i)]{facchinei:fischer:piccialli:07}. In addition, if the pair $(\bld x^*,\lambda^*)$ satisfies the KKT conditions in \eqref{eq:KKT_VI}, then $\bld x^*$ and the vectors $\lambda_1^*=\dots=\lambda_N^*=\lambda^*$ satisfy the KKT conditions for the GNE, i.e. \eqref{eq:KKT_game}  \cite[Th.~3.1(ii)]{facchinei:fischer:piccialli:07}. 

Note that under Standing Assumptions~\ref{ass:convex_constr_set}--\ref{ass:subgrad_lipschitz_strong_mon} the set of v-GNE is guaranteed to be a singleton \cite[Cor.~2.2.5;~Th.~2.3.3]{facchinei:pang}.

\section{Synchronous Distributed GNE Seeking}
\label{sec:synch_case}
In this section, we describe the \textit{\underline{S}ynchronous \underline{D}istributed \underline{G}N\underline{E} Seeking Algorithm with \underline{No}de variables} (SD-GENO). First, we outline the communication graph supporting the communication between agents, then we derive the algorithm via an operator splitting methodology.
 
\subsection{Communication network}
\label{sec:com_network}
The communication between agents is described by an \textit{undirected and connected} graph $\cal G =(\ca N,\ca E)$ where $\ca N $ is the set of players and $\ca E \subseteq \ca N \times\ca N$ is the set of edges. We define $|\ca E |= M$, and  $|\ca N |= N$.  If an agent $i$  shares information with $j$, then $(i,j)\in\ca E$, then we say that $j$ belongs to the neighbours of $i$, i.e., $j\in\ca N_i$ where $\ca N_i$ is the neighbourhood of $i$. %We define $|\ca E |= M$, and  $|\ca N |= N$.  If an agent $i$  shares information with another element of the network $j$, then $(i,j)\in\ca E$. We say that $j$ belongs to the neighbours of $i$, i.e., $j\in\ca N_i$ where $\ca N_i$ is the neighbourhood of $i$.% and define the associated edge as the couple $e_l:=(i,j)$ where $l\in\{1,\dots,M\}$.
%Assuming the graph to be undirected means that $(i,j)\in\ca E$ implies $(j,i)\in\ca E$. 
%A path between two agents $i$ and $h$ in $\ca N$ is an ordered sequence of elements of $\ca N$, starting in $i$ and ending in $h$, such that each couple of consecutive elements in the sequence belongs to $\ca E$. %, namely that for each element in the sequence the next one is in its neighbourhood.
 %A graph $\ca G$ is connected if $\forall i,j\in\ca N$, it exists a path from $i$ to $j$.
 Let us label the edges $e_l$, for $l\in\{1,\dots,M\}$. We denote by $E\in\bR^{M\times N}$ the \textit{incidence matrix}, where $[E]_{li}$ is equal to $1$ (respectively $-1$) if $e_l=(i,\cdot)$ ($e_l=(\cdot,i)$) and $0$ otherwise. By construction, $E\bld 1_N = \bld 0_N$. Then, we define   $\ca E_i^{\mathrm{out}}$ (respectively $\ca E_i^{\mathrm{in}}$) as the set of all the indexes $l$ of the edges $e_l$ that start from (end in) node $i$, moreover $\ca E_i=\ca E_i^{\mathrm{out}} \cup \ca E_i^{\mathrm{in}}$ .
%The unordered couples, i.e. edges, that compose $\ca E$ can be labelled as $e_l$, $l\in\{1,\dots,M\}$. Aiming to define the \textit{incidence matrix} $E\in\bR^{M\times N}$ of the graph $\ca G$, we arbitrarily order $e_l=(i,j)$ as an edge from $i$ to $j$. The element $[E]_{li}$ of $E$ is equal to $1$ (respectively $-1$) if $e_l=(i,\cdot)$ ($e_l=(\cdot,i)$) and $0$ otherwise. Notice that, by construction, $E\bld 1_N = \bld 0_N$. We define the set of all the edges $e_l$ starting  from (ending in) node $i$ as $\ca E_i^{\mathrm{out}}$ ($\ca E_i^{\mathrm{in}}$) and also $\ca E_i=\ca E_i^{\mathrm{out}} \cup \ca E_i^{\mathrm{in}}$ . 
The \textit{node Laplacian }$L\in\bR^{N\times N}$ of an undirected graph is a symmetric matrix and can be expressed as $L=E^\top E$,  \cite[Lem.~8.3.2]{Godsil:algebraic_graph_theory}. In the remainder of the paper, we exploit the fact that the Laplacian matrix is such that $ L\bld 1_N = \bld 0_N $ and $\bld 1^\top_N L = \bld 0^\top_N $. %This will allow us to decrease the number of auxiliary variables in ours algorithms.

\subsection{Algorithm design}
\label{sec:alg_develop_synch}

Now, we present a distributed algorithm with convergence guarantees to the unique v-GNE of the game in \eqref{eq:game_formulation}. The KKT system in \eqref{eq:KKT_game}, can be cast in compact form as
\begin{equation} \label{eq:compact_KKT_game}
\begin{split}
\bld 0 &\in F(\bld x)+\Lambda^\top \bld \lambda+N_{\bld \Omega}(\bld x) \\
\bld 0 &\in \bar b-\Lambda\bld x + N_{\bR^{mN}_{\geq 0}}(\bld \lambda)
\end{split} \:,
\end{equation}
where $\bld \lambda = \col(\lambda_1,\dots,\lambda_N)\in\bR^{mN}$, $\Lambda= \diag(A_1,\dots,A_N)\in\bR^{mN\times n}$ and $\bar b = \col (b_1,\dots,b_N)\in\bR^{mN}$. 

As highlighted before, for an agent $i$, a solution of its local optimality conditions is given by the strategy $x_i$ and the dual variable $\lambda_i$. To enforce consensus among the dual variables, hence obtain a v-GNE,  we introduce the auxiliary variables $\sigma_l,\, l\in\{1,\dots,M\}$, one for every edge of the graph. Defining $\bld \sigma=\col (\sigma_1,\dots,\sigma_M)\in\bR^{mM}$ and using $\bld E = E\otimes I_m \in \bR^{mM\times mN}$, we augment the inclusion in \eqref{eq:compact_KKT_game} as
\begin{equation} \label{eq:mod_compact_KKT_game}
\begin{split}
\bld 0 &\in F(\bld x)+\Lambda^\top \bld \lambda+N_{\bld \Omega}(\bld x) \\
\bld 0 &\in \bar b-\Lambda\bld x + N_{\bR^{mN}_{\geq 0}}(\bld \lambda) +\bld E^\top \bld \sigma\\
\bld 0 &\in -\bld  E \bld \lambda \:.
\end{split} 
\end{equation}
The variables $\{\sigma_l\}_{l\in \{1\dots M\}}$ are used to simplify the analysis, but we will show how we decrease their number to one for each node, increasing the scalability of the algorithm, especially for dense networks.

From an operator theoretic perspective, a solution $\varpi^* = \col (\bld x^*,\bld \sigma^*,\bld \lambda^*)$ to \eqref{eq:mod_compact_KKT_game} can be interpreted as a zero of the sum of two operators, $\ca A$ and $\ca B$, defined as
\begin{equation}
\label{eq:operators_def}
\begin{split}
\ca A : \varpi \mapsto &\begin{bmatrix}
0 & 0 & \Lambda^\top  \\
0 & 0 & -\bld E  \\
 -\Lambda & \bld E^\top & 0
\end{bmatrix}+\begin{bmatrix}
N_{\bld \Omega} (\bld x)\\
0\\
N_{\bR^{mN}_{\geq 0}} (\bld \lambda)
\end{bmatrix}\\
\ca B : \varpi \mapsto &\begin{bmatrix}
F(\bld x)\\
0\\
\bar b
\end{bmatrix}\,.
\end{split}
\end{equation} 
In fact, $\varpi^*\in\mathrm{zer}(\ca A+\ca B)$ if and only if $\varpi^*$ satisfies \eqref{eq:mod_compact_KKT_game}.

Next, we show that the zeros of $\ca A+\ca B$ are actually the v-GNE of the initial game. 
\smallskip
\begin{proposition}\label{prop:zer_AB_are_vGNE}
Let $\ca A$ and $\ca B$ be as in \eqref{eq:operators_def}. Then the following hold:
\begin{enumerate}[(i)]
\item $\mathrm{zer}(\ca A +\ca B)\not = \varnothing$ ,
\item if $ \col (\bld x^*,\bld \sigma^*,\bld \lambda^*)\in\mathrm{zer}(\ca A +\ca B)$ then $(\bld x^*,\lambda^*)$ satisfies the KKT conditions in \eqref{eq:KKT_VI}, hence $\bld x^*$ is the v-GNE for the game in \eqref{eq:game_formulation}.
 \hfill\QEDopen 
\end{enumerate} 
\end{proposition}
\smallskip
 The proof exploits the property of the incidence matrix $E$ of having the same null space of $L$, i.e. $\mathrm{ker}(E)=\mathrm{ker}(L)$, and the assumption that the graph is  connected. It can be obtained via an argument analogue to the one used in \cite[Th.~4.5]{pavel2017:distributed_primal-dual_alg}, hence we omit it.

%\subsection{Synchronous fully distributed algorithm} 
The problem of finding the zeros of the sum of two monotone operators is widely studied in literature and a plethora of different splitting method can be used to iteratively solve the problem \cite{eckstein1989:splitting}, \cite[Ch.~26]{bauschke2011convex}.  
A necessary first step is to prove the monotonicity of the defined operators.
\smallskip
\begin{lemma}\label{lem:max_mon_of_operators}
The mappings $\ca A$ and $\ca B$ in \eqref{eq:operators_def} are maximally monotone. Moreover, $\ca B $ is  $\frac{\alpha}{\ell^2}$-cocoercive. \hfill\QEDopen
\end{lemma}
\smallskip
The splitting method chosen here to find $\mathrm{zer}(\ca A+\ca B)$ is the \textit{preconditioned forward-backward} splitting (PFB), which can be applied thanks to the properties stated in Lemma~\ref{lem:max_mon_of_operators}. 
\smallskip
\begin{remark}
The choice is driven by two main features simplicity and implementability. In fact, the PFB requires only one round of communication between agents at each iteration, minimizing in this way the most demanding operation in multi-agent algorithms, i.e., information sharing.
% For this reason, more complex splitting like the Tseng's Splitting \cite[Ch.~28.6]{bauschke2011convex} or the Douglas-Rachford Splitting \cite[Ch.~28.3]{bauschke2011convex} were discarded. During the choice also a simpler algorithm, the proximal point algorithm \cite[Prop.~23.38]{bauschke2011convex}, the implementation imposes to solve another game at each time step, therefore losing all the benefits of this approach.
\end{remark}
\smallskip
The iteration of the algorithm takes the form of the so called Krasnosel'ski\u i iteration, namely 
\begin{equation}
\label{eq:Krasno_iter_synch}
\begin{split}
\tilde\varpi^k &= T \varpi^k \\
\varpi^{k+1}&=\varpi^k+\eta (\tilde\varpi^k-\varpi^k)
\end{split}
\end{equation} 
where $\varpi^k = \col(\bld x^k,\bld\sigma^k,\bld\lambda^k)$, $\eta>0$ and $T$ is the PFB splitting operator
\begin{equation}
\label{eq:T_PFB_operator}
T = \mathrm{J}_{\gamma\Phi^{-1}\ca A}\circ(\Id-\gamma\Phi^{-1}\ca B)\, ,
\end{equation}
 where $\gamma>0$ is a step size. The so-called preconditioning matrix $\Phi$ is defined as
 \begin{equation}
 \label{eq:preconditioning_matrix}
 \Phi:=\begin{bmatrix}
	\bld \tau^{-1}   & 0  & -\Lambda^\top\\
	0 & \delta^{-1}I_{mM}  & \bld E \\
	-\Lambda & \bld E^\top & \bld \varepsilon^{-1}
 \end{bmatrix}
\end{equation}   
where $\delta\in\bR_{>0}$,  $\bld \varepsilon = \diag(\varepsilon_1,\dots,\varepsilon_N)\otimes I_m $ with $\varepsilon_i>0,\:\forall i\in \ca N$ and $\bld \tau$ is defined in a similar way.

From \eqref{eq:T_PFB_operator}, we note that $\fix(T)=\mathrm{zer}(\ca A + \ca B)$, indeed $\varpi \in\fix(T) \iffs \varpi \in T \varpi \iffs 0 \in \Phi^{-1}(\ca A+\ca B)\varpi \iffs \varpi\in\zer(\ca A +\ca B)$, \cite[Th.~26.14]{bauschke2011convex}.  Thus, the zero-finding problem is translated into the fixed point problem for the mapping $T$ in \eqref{eq:T_PFB_operator}.
 
At this point, we calculate from \eqref{eq:Krasno_iter_synch} the explicit update rules of the variables. We first focus on the first part of the update, i.e., $\tilde \varpi^k=T\varpi^k$. It can be rewritten as $\tilde \varpi^k \in \mathrm{J}_{\gamma\Phi^{-1}\ca A}\circ(\Id-\gamma\Phi^{-1}\ca B)\varpi^k \iffs \Phi(\varpi^k -\tilde\varpi^k) \in \ca A \tilde \varpi^k +\ca B \varpi^k$ and finally 
\begin{equation}\label{eq:inclusion_synch}
\bld 0\in  \ca A \tilde \varpi^k +\ca B \varpi^k + \Phi(\tilde\varpi^k -\varpi^k)\:,
\end{equation}
 here $\tilde \varpi^k :=\col (\tilde {\bld x}^k,\tilde {\bld\sigma}^k, \tilde {\bld\lambda}^k)$. For ease of notation, we drop the time superscript $k$. By solving the first row block of \eqref{eq:inclusion_synch}, i.e.
 $\bld 0 \in F(\bld x) +N_{\bld \Omega}(\tilde{\bld x}) + \bld \tau^{-1}(\tilde{\bld x} -\bld x) + \Lambda^\top\bld \lambda$, we obtain
\begin{equation}\label{eq:row3_sync}
\tilde{\bld x} = \mathrm{J}_{N_{\bld \Omega}} \circ\big(\bld x-\bld \tau(F(\bld x)+\Lambda^\top \bld \lambda ) \big)\,.
\end{equation} 
The third row block of \eqref{eq:inclusion_synch} instead reads as $\bld 0 \in \bar b +N_{\bR^{mN}_{\geq 0}}(\tilde{\bld \lambda}) +\Lambda(2\tilde{\bld x}-\bld x) + \bld E^\top(2\tilde{\bld \sigma}-\bld \sigma)+ \bld \varepsilon^{-1}(\tilde{\bld\lambda}-\bld\lambda)$ that leads to 
\begin{equation}\label{eq:row2_sync}
\tilde{\bld \lambda} = \mathrm{J}_{N_{\bR^{mN}_{\geq 0}}}\circ \big(\bld \lambda-\bld \varepsilon( \Lambda (2\tilde{\bld x}-\bld x) -\bar b - \bld E^\top(2\tilde{\bld \sigma}-\bld\sigma )) \big)\,.
\end{equation} 
 The second row block of \eqref{eq:inclusion_synch} defines the simple update $\tilde{\bld \sigma} = \bld \sigma +  \delta \bld E\bld \lambda $. We note that in the update \eqref{eq:row2_sync} of $\tilde{\bld \lambda}$, only $\bld E^\top\bld \sigma $ is used, hence an agent $i$ needs only an aggregated information over the edge variables $\{\sigma_l\}_{l\in\ca E_i}$, to  update its state and  the dual variables. We exploit this property by replacing the edge variables with $\bld z = \bld E^\top \bld \sigma\in\bR^{Nm}$. In this way, the auxiliary variables are one for each agent, instead of being one for each edge. Using the property $\bld E^\top\bld E =L\otimes I_m=\bld L$, we cast the update rule of these new auxiliary variables as
 \begin{equation}\label{eq:row1_mod_sync}
 \begin{split}
&\tilde{\bld z}^k = \bld z^k + \delta \bld L \bld \lambda^k  \\
&\bld z^{k+1} = \bld z^k  +\eta (\tilde{\bld z}^k -\bld z^k) \, .
 \end{split}
\end{equation}  
By introducing  $\bld z$ in \eqref{eq:row2_sync}, we then have
\begin{equation}\label{eq:row2_sync_mod}
\tilde{\bld \lambda} = \mathrm{J}_{N_{\bR^{mN}_{\geq 0}}}\circ \big(\bld \lambda+\bld \varepsilon( \Lambda ( 2\tilde{\bld x}-\bld x) -\bar b -2\tilde{\bld z} +\bld z  ) \big)\,.
\end{equation}
%%\smallskip
%\begin{remark}
%If the initial value of $z$ is such that $\bld{1}^T_{mN} \bld z_0 = 0$ then from \eqref{eq:row1_mod_sync} it holds that $\bld{1}^T_{mN} \bld z_k= 0$,  for every time instant $k$.
%\end{remark}
The next theorem shows that an equilibrium of the new mapping is a v-GNE.
\smallskip
\begin{theorem}\label{th:eq_are_vGNE_mod_map}
If $\col(\bld x^*, \bld z^*, \bld \lambda^* )$ is a solution to the equations \eqref{eq:row3_sync}, \eqref{eq:row1_mod_sync} and  \eqref{eq:row2_sync_mod}, with $\bld 1^\top\bld z^*=0$, then $\bld x^*$ is a v-GNE.   
\hfill\QEDopen 
\end{theorem}
\smallskip
\begin{remark}
The change of auxiliary variables, from $\bld \sigma$ to $\bld z$, is particularly useful in large non-so-sparse networks and it is in general convenient when the number of edges higher than the number of nodes. In fact, for dense networks, we have one auxiliary variable for each player, hence the scalability of the algorithm is preserved.   
\end{remark}
\subsection{Synchronous, distributed algorithm with node variables (SD-GENO)} 
\begin{algorithm}[t]
\DontPrintSemicolon
\textbf{Input:} $k=0$, $\bld x^0 \in \bR^{n} $, $\bld \lambda^0 \in\bR^{mN}$, $\bld z^0=\bld 0_{mN}$, and chose $\eta,\, \delta,\, \bld \varepsilon,\, \bld \tau$ as in Theorem~\ref{th:convergence_sync}. \;
\For{$i\in\ca N$}{
$\tilde{ x}_{i}^k=\mathrm{proj}_{\Omega_i} \big( x_{i}^k-\tau_i(\nabla_i f_i(x_{i}^k,\bld x_{-i}^k)+ A_i^\top \lambda_{i}^k ) \big) $ \;
$\tilde{z}_{i}^k = z_{i}^k +\delta \sum_{j\in\ca N_i }  (\lambda_{i}^k - \lambda_{j}^k) $\;
$\tilde{\lambda}_{i}^k  = \mathrm{proj}_{\bR^{m}_{\geq 0}} \big( \lambda_{i}^k+\varepsilon_i\left( A_i(2\tilde{x}_{i}^k - x_{i}^k) \right. $\;
$\hspace{4cm }\left. - b_i+z_{i}^k -2\tilde{z}_{i}^k  \right) \big)$\;
$x_{i}^{k+1} =  x_{i}^k +\eta(\tilde{ x}_{i}^k - x_{i}^k)$\;
$z_{i}^{k+1} =  z_{i}^k +\eta(\tilde{ z}_{i}^k -z_{i}^k)$ \;
$\lambda_{i}^{k+1} = \lambda_{i}^k +\eta(\tilde{ \lambda}_{i}^k - \lambda_{i}^k)$\;
$k\leftarrow k+1$\;
}
\caption{SD-GENO}
\label{alg:synch_alg}
\end{algorithm}
%In this section we describe the local update 
We are now ready to state the update rules defining the synchronous version of the proposed algorithm. 
The update rule is obtained by gathering \eqref{eq:row3_sync}, \eqref{eq:row1_mod_sync}, \eqref{eq:row2_sync_mod} and by modifying the second part of \eqref{eq:Krasno_iter_synch} via the auxiliary variables $\bld z$:
\begin{equation}\label{eq:update_compact_complete_alg}
\begin{split}
&\tilde{\bld x}^{k}  = \mathrm{proj}_{\bld \Omega} \big(\bld x^k-\bld \tau(F(\bld x^k)+\Lambda^\top \bld \lambda^k ) \big) \\
&\tilde{\bld z}^{k} = \bld z^k + \delta \bld L \bld \lambda^k\\
&\tilde{\bld \lambda}^{k}  = \mathrm{proj}_{\bR^{mN}_{\geq 0}} \big(\bld \lambda^k+\bld \varepsilon( \Lambda (2\tilde{\bld x}^k-\bld x^k) -\bar b -2\tilde{\bld z}^k +\bld z^k ) \big)\\
&\bld x^{k+1} = \bld x^k +\eta(\tilde{\bld x}^k -\bld x^k)\\
&\bld z^{k+1} = \bld z^k +\eta(\tilde{\bld z}^k -\bld z^k)\\
&\bld \lambda^{k+1} = \bld \lambda^k +\eta(\tilde{\bld \lambda}^k -\bld \lambda^k) \:,\\
\end{split}
\end{equation}  
See also Algorithm~\ref{alg:synch_alg}, for the local updates.

%where we used the fact that  $\mathrm{J}_S=\mathrm{proj}_S$ for a convex nonempty and closed set $S$ \cite[Ex.~23.4]{bauschke2011convex}. 
%
%The final formulation of SD-GENO is obtained from \eqref{eq:update_compact_complete_alg} specifying the local update rule for each player of the game.
The convergence of Algorithm~\ref{alg:synch_alg} to the v-GNE of the game in \eqref{eq:game_formulation} is guaranteed by the following theorem.
\smallskip
\begin{theorem}\label{th:convergence_sync}
Let $\vartheta>\frac{\ell^2}{2\alpha}$,  $\bld \varepsilon$, $ \delta$, $\bld \tau>0$ such that $\Phi-\vartheta I\succeq 0$ and  $\eta\in(0,\frac{4\alpha\vartheta-\ell^2}{2\alpha\vartheta})$. Then, Algorithm~\ref{alg:synch_alg} converges to the v-GNE of the game in \eqref{eq:game_formulation}.    
\hfill\QEDopen
\end{theorem}
\smallskip
 
\section{Asynchronous Distributed Algorithm }
\label{sec:asynch_case}
In this section, we present the main contribution of the paper, the  \textit{\underline{A}synchronous \underline{D}istributed \underline{G}N\underline{E} Seeking Algorithm with \underline{No}de variables} (AD-GENO), namely, the asynchronous counterpart of Algorithm~\ref{alg:synch_alg}. As in the previous section, we first define a preliminary version of the algorithm using the edge auxiliary variables $\bld \sigma$, and then we derive the final formulation via the variable $\bld z$. To achieve an asynchronous update of the agent variables, we adopt the ``ARock" framework \cite{peng2016arock}.% This allows us to model not only the asynchronicity, but also the delays in the communication between players. 
\smallskip
\subsection{Algorithm design}
We modify the update rule in \eqref{eq:Krasno_iter_synch} to describe the asynchronism, in the local update of the agent $i$, as follows
%: the case in which only the variables of an agent $i$ are updated at time $k$ and the other stay the same. The update rules reads as
\begin{equation}\label{eq:Krasno_asynch}
\varpi^{k+1}=\varpi^k+\eta \Upsilon_i(T\varpi^k-\varpi^k)\,,
\end{equation}
%where $\Upsilon_i\in\bR^D$ is a diagonal matrix. We define  $[\Upsilon_{i}]_j$ equal to $1$, if the $j$-th element of $\bld \varpi$ is an element of $\col(x_i,\{\sigma_l\}_{l\in\ca E^{\mathrm{out}}_i},\lambda_i)$ and $0$ otherwise. So, the update of the edge variable $\sigma_l$ is taken care by agent $i$ iff $l\in\ca E^{\mathrm{out}}_i$. 
where $\Upsilon_i$ is a real diagonal matrix of dimension $n+(N+M)m$, where the element $[\Upsilon_{i}]_{jj}$ is $1$ if the $j$-th element of $\col( \bld x,\: \bld\sigma,\:\bld \lambda)$ is an element of $\col(x_i,\:\{\sigma_l\}_{l\in\ca E ^{\mathrm{out}}_i},\: \lambda_i)$ and $0$ otherwise.% (see also \cite{peng:2016:coord_friendly}).
%
% Define $\Upsilon_i:=\mathrm{blkdiag} (\Upsilon_{x,i}, \Upsilon_{\sigma,i}, \Upsilon_{\lambda,i})$ where $\Upsilon_{\lambda,i}\in\bR^{nN}$,  $\Upsilon_{\sigma,i}\in\bR^{mM}$ and  $\Upsilon_{\lambda,i}\in\bR^{mN}$  are diagonal matrices. The element $[\Upsilon_{x,i}]_{jj}$ ( $[\Upsilon_{\lambda,i}]_{jj}$) is $1$ if the $j$-th element of $\bld x$ ($\bld \lambda$) is an element of $x_i$ ($\lambda_i$) and $0$ otherwise. Instead, $[\Upsilon_{\sigma,i}]_{jj}$ is equal to $\textstyle{\frac{1}{2}}$ if  the $j$-th element of $\bld \sigma$ is a component of $\sigma_i$ and $0$ otherwise.
%By using this definition, we obtain a set $\bld \Upsilon:=\{ \Upsilon_i \}_{i\in\ca N}$, that satisfies the requirement of \cite[Sec.~1.3]{peng2016arock}, i.e., $\sum_{i\in\ca N} \Upsilon_i = I_D$ and $\sum_{i\in\ca N} \lVert \Upsilon_i x\rVert^2 \leq C \lVert x \rVert^2$, for some  $C>0$ and $\forall x\in\bR^{N(n+m)+mM}$ (see also \cite{peng:2016:coord_friendly}).
We assume that the choice of which agent performs the update at iteration $k\in\bN_{\geq 0}$ is ruled by an i.i.d. random  variable $\zeta^k$, that takes values in $\bld \Upsilon:=\{ \Upsilon_i \}_{i\in\ca N}$. Given a discrete probability distribution $(p_1,\dots,p_N)$, let $\mathbb{P}[\zeta^k = \Upsilon_i] = p_i$, $\forall i\in \ca N$. Therefore, the formulation in \eqref{eq:Krasno_asynch} becomes 
\begin{equation}\label{eq:Krasno_asynch_zeta}
\varpi^{k+1}=\varpi^k+\eta \zeta^k(T\varpi^k-\varpi^k)\,.
\end{equation}

We also consider the possibility of delayed information, namely the update \eqref{eq:Krasno_asynch_zeta} can be performed with outdated values of $\varpi^k$. We refer to \cite[Sec.~1]{peng2016arock} for a more complete overview on the topic. Due to the structure of the $\Upsilon_i$, the update of $x_i$, $\lambda_i$ and $\{\sigma_l\}_{l\in\ca E ^{\mathrm{out}}_i} $ are performed at the same moment, hence they share the same delay $\varphi_i^k$ at $k$. %On the other hand, $\{\sigma_l\}_{l\in\ca E_{i}}$ is updated by both the agents that the edge connects. Therefore, if  $l\in\ca E^{\mathrm{out}}_i$ and $l\in\ca E^{\mathrm{in}}_j$, the delay of $\sigma_l$ at time $k$ is $\psi^k_l:=\min(\varphi^k_i,\,\varphi^k_j  )$. 

We denote the vector of possibly delayed information at time $k$ as $\hat\varpi^k$, hence the reformulation of  \eqref{eq:Krasno_asynch_zeta} reads as
\begin{equation}\label{eq:Krasno_asynch_final}
\varpi^{k+1}=\varpi^k+\eta \zeta^k(T-\Id)\hat\varpi^k\,.
\end{equation} 

Now, we impose that the maximum delay is uniformly bounded. 
\smallskip
\begin{stassumption}[Bounded maximum delay]\label{ass:bounded_delay}
The delays are uniformly upper bounded, i.e. there exists $\bar\varphi>0$ such that $\sup_{k\in\bN_{\geq 0}}\max_{i\in\ca N}\{ \varphi_i^k \}\leq \bar\varphi<+\infty$. \hfill\QEDopen
\end{stassumption} 
\smallskip
%\begin{remark}
%During the update of an agent $i$, the agent stores an unchanging copy of $\hat \varpi_k$ that is used in the operation, hence if another player finishes its update in the meantime the changing in its relative variables will not effect the current computation of player $i$.
%\end{remark}
%\smallskip

From the computational perspective, we assume that each player $i$ has a public and a private memory. The first stores the information obtained by the neighbours $\ca N_i$. The private is instead used during the update of $i$ at time $k$ and it is an unchangeable copy of the public memory at iteration $k$.  The local update rules in Algorithm~\ref{alg:E-ADAGNES} are obtained similarly to Sec.~\ref{sec:alg_develop_synch} for SD-GENO, hence by using the definition of $T$. The obtained algorithm resembles ADAGNES in \cite[Alg.~1]{Pavel:Yi:2018:Asynch}, therefore we name it E-ADAGNES.
%At each iteration $k$ the update of agent $i_k$, obtained by mean of $\zeta_k$, is composed of three parts: 
\begin{algorithm}[h!]
\DontPrintSemicolon
\textbf{Input:} $k=0$, $\bld x^0 \in \bR^{n} $, $\bld \lambda^0 \in\bR^{mN}$, $\bld \sigma^0=\bld 0_{mM}$, chose $\eta,\, \delta,\, \bld \varepsilon,\, \bld \tau$ as in Theorem~\ref{th:convergence_sync}.  \;
\hrule
\smallskip
\textbf{Iteration $k$:} Select the agent $i^k$ with probability $\mathbb{P}(\zeta^k=\Upsilon_{i^k})=p_{i^k}$\; 
\textbf{Reading:} Agent $i^k$ copies in its private memory the current values of the public memory, i.e. $x_{j}^{k-\varphi^k_j}$, $\lambda_{j}^{k-\varphi^k_j}$ for $j\in\ca N_{i^k}$ and $\sigma_{l}^{k-\psi^k_l}, \: \forall l\in\ca E_{i^k}$ \; 
\textbf{Update:}\;
$\tilde{ x}_{i^k}^k  = \mathrm{proj}_{\Omega_{i^k}} \big( x_{i^k}^{k-\varphi^k_{i^k}}-\tau_{i^k}(\nabla_{i^k} f_{i^k}(x_{i^k}^{k-\varphi^k_{i^k}},\hat{\bld x}_{-i^k}^k) + A_{i^k}^\top \lambda_{i^k}^{k-\varphi^k_{i^k}} ) \big) $ \;\bigskip
$\tilde{\sigma}_{l}^k = \sigma_{l}^{k-\varphi^k_{i_k}} + \delta ([E]_{l}\otimes I_m )\hat{\bld\lambda}^{k} \:,\quad \forall l\in \ca E_{i^k}^{\mathrm{out}}$\;\bigskip
$\tilde{\lambda}_{i^k}^k  = \mathrm{proj}_{\bR^{m}_{\geq 0}} \big( \lambda_{i^k}^{k-\varphi^k_{i^k}}+\varepsilon_{i^k}( A_{i^k}(2\tilde{x}_{i^k}^k - x_{i^k}^{k-\varphi^k_{i^k}}) - b_{i^k} - ([E^\top]_{i^k}\otimes I_m)(2 \tilde{\bld \sigma}^{k}  - \hat{\bld \sigma}^{k}   )) \big)$\;\bigskip
$x_{i}^{k+1} =  x_{i}^{k-\varphi^k_{i^k}} +\eta(\tilde{ x}_{i}^k - x_{i}^k-\varphi^k_{i^k})$\;
$\sigma_{l}^{k+1} =  \sigma_{l}^{k-\varphi^k_{i_k}} +\eta(\tilde{ \sigma}_{l}^k -\sigma_{l}^{k-\psi^k_l})\:,\quad \forall l\in \ca E_{i^k} $ \;
$\lambda_{i}^{k+1} = \lambda_{i}^{k-\varphi^k_{i^k}} +\eta(\tilde{ \lambda}_{i}^k - \lambda_{i}^{k-\varphi^k_{i^k}})$\;
\textbf{Writing:} Agent $i^k$ writes in the public memories of  $j\in\ca N_{i^k}$  the new values of  $x_{i^k}^{k+1}$, $\lambda_{i^k}^{k+1}$ and $\{\sigma_{l}^{k+1}\}_{l\in\ca E_{i^k}^{\mathrm{out}}}$\; \smallskip
 $k\leftarrow k+1$\;
\caption{E-ADAGNES}
\label{alg:E-ADAGNES}
\end{algorithm}
%
%
%\begin{enumerate}[(i)]
%\item \textit{Reading:} agent $i^k$ copies its public memory at time $k$ into its private memory.
%\item \textit{Computation:} the player updates $x_{i^k}$, $\lambda_{i^k}$ and $\{\sigma_l\}_{l\in\ca E_{i_k}}$ via \eqref{eq:Krasno_asynch_final}.
%\item \textit{Writing:} the agent writes in the public memories of  $j\in\ca N_{i_k}$  the new values of  $x_{i_k}$, $\lambda_{i_k}$ and $\{\sigma_l\}_{l\in\ca E_{i_k}\cap \ca E_{j}}$.
%\end{enumerate}
%The local update rules for the computing step are obtained as done in Sec.~\ref{sec:alg_develop_synch}, by exploiting the structure of $T$. 
%\smallskip
%\begin{remark}
%The obtained algorithm resemble the structure of ADAGNES in \cite[Algorithm~1]{Yi:Pavel:2018:Asynch}, therefore we refer to it for more complete description, we will instead focus more on the counterpart that uses only node variables to reach convergence.
%\end{remark}
%\smallskip

The convergence of the update \eqref{eq:Krasno_asynch_final} is proven relying on the theoretical results provided in \cite{peng2016arock} for the Krasnosel'ski\u i asynchronous iteration.
\begin{theorem}\label{th:convergence_asynch_sigma} 
Let $\eta\in(0,\frac{4\alpha\vartheta-\ell^2}{\alpha\vartheta}\frac{cNp_{\min}}{4\bar\varphi\sqrt{p_{\min}}+1}]$, where $p_{\min}:=\min \{p_i\}_{i\in\ca N}$ and $c\in(0,1)$. Then, the sequence $\{\bld x^k\}_{k\in\bN_{\geq 0}}$ defined by Algorithm~\ref{alg:E-ADAGNES} converges to the v-GNE of the game in \eqref{eq:game_formulation} almost surely.  \hfill\QEDopen
\end{theorem}

\subsection{Asynchronous, distributed algorithm with node variables (AD-GENO)}
We complete the technical part of the paper by performing the change from auxiliary variables over the edges to variables over the nodes, attaining in this way the final formulation of our proposed algorithm. With Algorithm~\ref{alg:E-ADAGNES} as starting point, we show that this change does not affect the dynamics of the pair $(\bld x,\bld \lambda)$, thus preserving the convergence.

However, in this case, we need to introduce an extra variable for each node $i$, i.e., $\mu_i\in\bR^{m}$. This is an aggregate information that groups all the changes of the neighbours dual variables from the previous update of $i$ to the present iteration. We highlight that these variables are updated during the writing phase of the neighbours, therefore they do not require extra communications between the agents. 
\smallskip
\begin{remark}
The need for $\mu_i\in\bR^{m}$ arises from the different update frequency between $\{\sigma_l\}_{l\in\ca E_i}$ and $ z_i$. % While the latter is updated exclusively by $i$, the former  is instead updated by both $i$ and $j$. 
 Therefore, we cannot characterize the dynamics of $\bld \sigma$, if we define $\bld z=\bld E^\top\bld \sigma$ only.  
\end{remark}
\smallskip
Algorithm~\ref{alg:AD-GENO} presents AD-GENO, where $\mu_i$ are rigorously defined.
\smallskip
\begin{algorithm}[h!]
\DontPrintSemicolon
\textbf{Input:} $k=0$, $\bld x^0 \in \bR^{n} $, $\bld \lambda^0 \in\bR^{mN}$, $\bld z^0=\bld 0_{mN}$, chose $\eta,\, \delta,\, \bld \varepsilon,\, \bld \tau$ as in Theorem~\ref{th:convergence_sync}. For all $i\in\ca N$ and $\mu_i=\bld 0_m$. \;
\hrule
\smallskip
\textbf{Iteration $k$:} Select the agent $i^k$ with probability $\mathbb{P}[\zeta^k=\Upsilon_{i^k}]=p_{i^k}$\; 
\textbf{Reading:} Agent $i^k$ copies in its private memory the actual values of the public memory, i.e. $x_{j}^{k-\varphi^k_j}$, $\lambda_{j}^{k-\varphi^k_j}$, $z_{j}^{k-\varphi^k_j}$ for $j\in\ca N_{i^k}$ and $\mu_{i}$. Reset the public values of $\mu_{i}$ to  $\bld 0_m$.\; 
\textbf{Update:}\;
$\tilde{ x}_{i^k}^k  = \mathrm{proj}_{ \Omega_{i^k}} \big( x_{i^k}^{k-\varphi^k_{i^k}}-\tau_{i^k}(\nabla_{i^k}  f_{i^k}(x_{i^k}^{k-\varphi^k_{i^k}},\hat{\bld x}_{-i^k}^{k})+ A_{i^k}^\top \lambda_{i^k}^{k-\varphi^k_{i^k}} ) \big) $ \;\smallskip
$\tilde{z}_{i^k}^k = z_{i^k}^{k-\varphi^k_i} +\delta\eta\mu_{i^k}  $\;\smallskip
$\tilde{\lambda}_{i^k}^k  = \mathrm{proj}_{\bR^{m}_{\geq 0}} \left( \lambda_{i^k}^{k-\varphi^k_{i^k}}+\varepsilon_{i^k}( A_{i^k}(2\tilde{x}_{i^k}^k - x_{i^k}^{k-\varphi^k_{i^k}}) \right. - b_{i^k}-\tilde z_{i^k}^{k-\varphi^k_i}\left. -2\delta\sum_{j\in\ca N_i\setminus \{i\}} (\lambda_{i^k}^{k-\varphi^k_{i^k}}-\lambda_{j}^{k-\varphi^k_{j}}) \right)$\;\bigskip
$x_{i}^{k+1} =  x_{i^k}^{k-\varphi^k_{i^k}} +\eta(\tilde{ x}_{i^k}^k - x_{i^k}^{k-\varphi^k_{i^k}})$\;
$z_{i^k}^{k+1} =  \tilde z_{i^k}^{k} + \eta\delta\sum_{l\in\ca E_{i_k}^{\mathrm{out}}} ([E]_l\otimes I_m )\hat{\bld \lambda}^k $ \;
$\lambda_{i^k}^{k+1} = \lambda_{i^k}^{k-\varphi^k_{i^k}} +\eta(\tilde{ \lambda}_{i^k}^k - \lambda_{i^k}^{k-\varphi^k_{i^k}})$\;
\textbf{Writing:} Agent $i^k$ writes in the public memories of  $j\in\ca N_{i^k}$  the new values of  $x_{i^k}^{k+1}$ and  $\lambda_{i^k}^{k+1}$, for $j\in\ca N_{i^k}\setminus\{i^k\}$ the player $i^k$ also overwrites $\mu_{j}$ as\;
 $\mu_{j} \leftarrow \mu_{j} + \lambda_{j}^{k-\varphi^k_{j}} - \lambda_{i}^{k-\varphi^k_{i^k}}$\;\smallskip
 %where $\mathrm{sgn}(\ca E)$ is $1$ if the edge connecting $i^k$ to $j$ belongs to $\ca E_i^{\mathrm{out}}$ and $-1$ otherwise.\;
 $k\leftarrow k+1$\;
\caption{AD-GENO}
\label{alg:AD-GENO}
\end{algorithm}
\smallskip

The convergence of AD-GENO is proven by the following theorem. Essentially, we show that introducing $\bld z$ does not change the dynamics of $(\bld x,\bld \lambda)$. 
\smallskip
\begin{theorem}\label{th:convergence_AD-GENO}
Let $\eta\in(0,\frac{4\alpha\vartheta-\ell^2}{\alpha\vartheta}\frac{cNp_{\min}}{4\bar\varphi\sqrt{p_{\min}}+1}]$ with $p_{\min}:=\min \{p_i\}_{i\in\ca N}$ and $c\in(0,1)$. Then, the sequence $\{\bld x^k\}_{k\in\bN_{\geq 0}}$ defined by Algorithm~\ref{alg:AD-GENO} converges to the v-GNE of the game in \eqref{eq:game_formulation} almost surely.  \hfill\QEDopen
\end{theorem}

\section{Simulation}
\label{sec:simulations}
This section presents the implementation of  AD-GENO to solve a network Cournot game, that models the interaction of $N$ companies competing over $m$ markets. The problem is widely studied and we adopt a set-up similar to the one in \cite{yu:2017distributed,pavel2017:distributed_primal-dual_alg}. We chose $N=8$ companies, each operating $4$ strategies, i.e., $x_i\in\bR^4$, $\forall i\in\ca N$. It ranges in $0\leq x_i \leq \Omega_i$, where $\Omega_i\in\bR^{4}$ ans its elements are  randomly drawn from $[10,45]$. The markets are $ m = 4 $, named $A$, $B$, $C$ and $D$. In Figure~\ref{fig:markets_and_network},  an edge between a company and a market is drawn, if at least one of that player's strategies is applied to that market. Two companies are neighbors if they share a  market.      
\begin{figure}
\centering
\subfloat[][]
{\includegraphics[scale = 0.15]{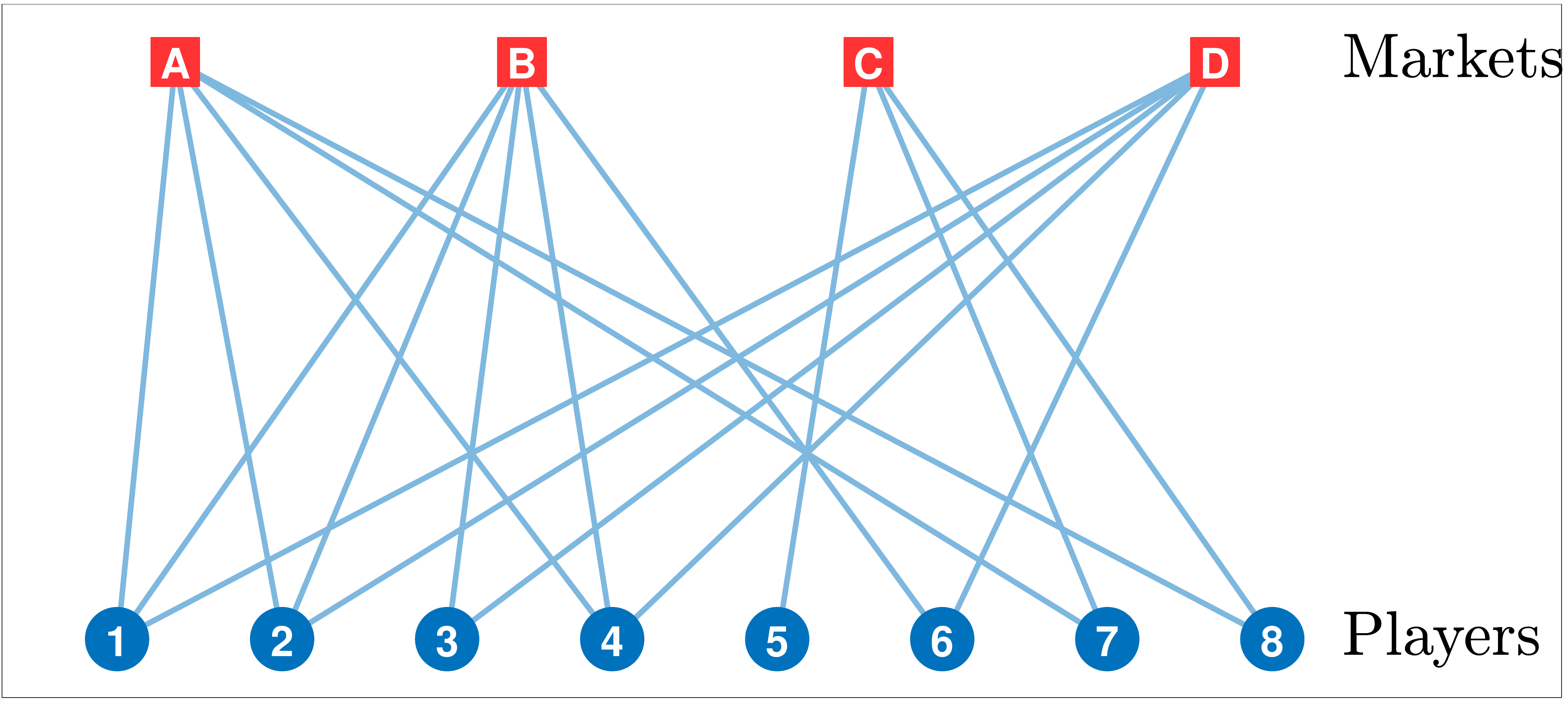}\label{fig:Markets}} \quad %[width=.225\textwidth] % How it was
\subfloat[][]
{\includegraphics[scale = 0.15]{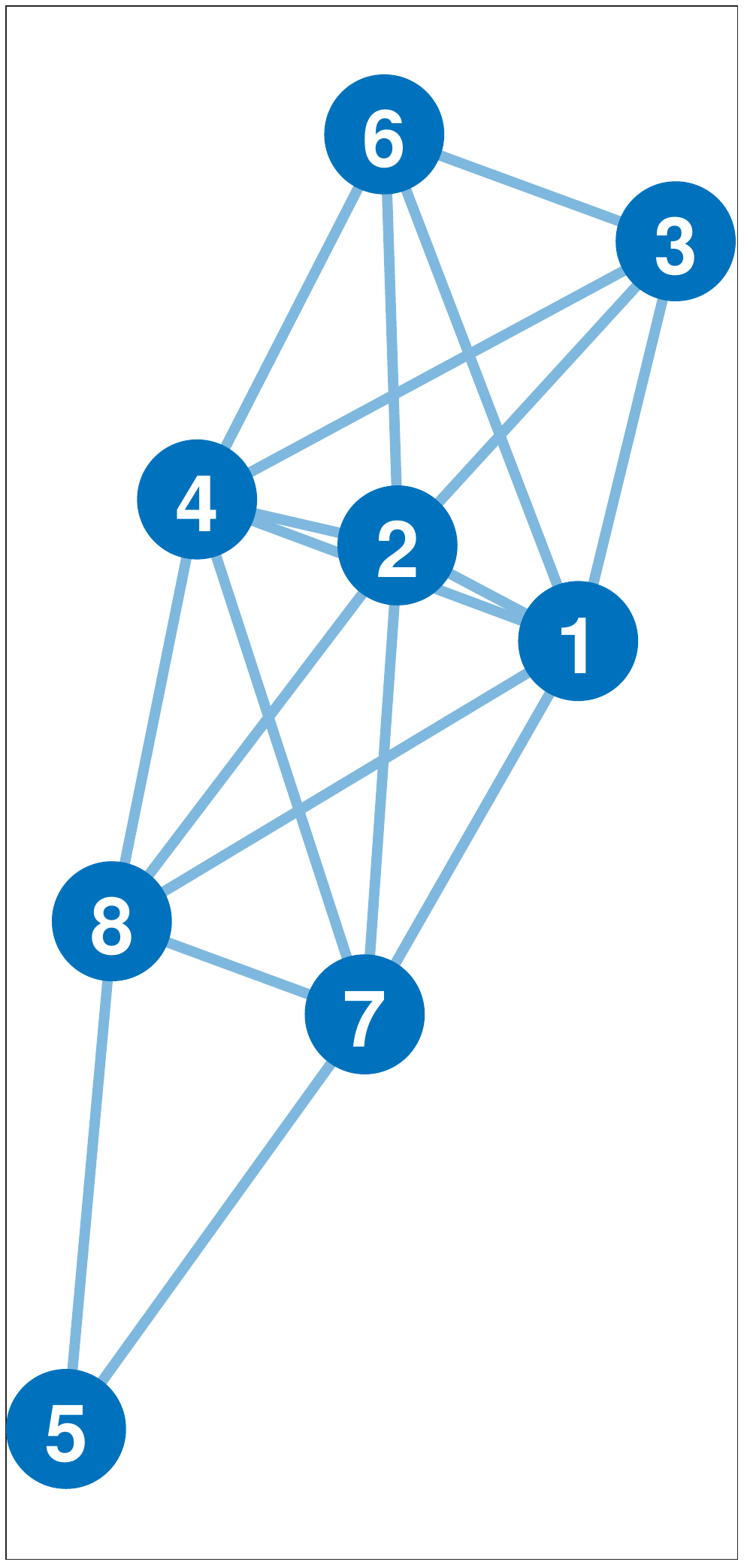}\label{fig:Network}} \\
\caption{(a) Interactions of the players $\{1,\dots,8\}$ with the markets $A$, $B$, $C$, $D$, (b) Communication network between players arising from the competition.}
\label{fig:markets_and_network}
\end{figure}
The constraint matrix is $\bld A=[A_1,\dots,A_N]\in \bR^{4\times 32}$ and the columns $k$ of $A_i$ have a nonzero element in position $j$ if the $k$-th strategy of player $i$ is applied to market $j$. The nonzero values are randomly chosen from $[0.6,1]$. The elements of $b\in\bR^{4}$ are the markets' maximal capacities and are randomly chosen from $[20,100]$. The arising inequality coupling constraint is $\bld{Ax}\leq b$. The local cost function is $f_i(x_i,\bld x_{-i})=c_i(\bld x) -P(\bld x)^\top A_ix_i$, where $c_i(\bld x)$ is the cost of playing a certain strategy and $P(\bld x)$ the price obtained by the market. We define the markets price as a linear function $P(\bld x)=\bar P -DA\bld x$, where $\bar P\in\bR^4 $ and $D\in\bR^{4\times 4}$ is a diagonal matrix, the values of their elements are randomly chosen respectively from $[250,500]$ and $[1,5]$. The cost function is quadratic $c_i(\bld x) = x_i^\top Q_i x_i + q_i^\top x_i$, where the elements of the diagonal matrix $Q_i\in\bR^{4\times 4}$ and the vector $q_i\in\bR^4$ are randomly drawn respectively from $[1,8]$ and $[1,4]$.   
%\begin{figure}
%\centering
%\subfloat[][]
%{\includegraphics[width=.22\textwidth]{Images/AD_GENO/XValue.pdf}\label{fig:Xval}} \quad %[width=.225\textwidth] % How it was
%\subfloat[][]
%{\includegraphics[width=.22\textwidth]{Images/AD_GENO/LambdaError.pdf}\label{fig:LambdaErr}} \\
%\subfloat[][]
%{\includegraphics[width=.22\textwidth]{Images/AD_GENO/LambdaVal.pdf}\label{fig:LambdaVal}} \quad
%\subfloat[][]
%{\includegraphics[width=.22\textwidth]{Images/AD_GENO/ConstrFeas.pdf}\label{fig:Constr}
%\label{fig:1_d}}
%\caption{(a)-(c) Trajectories of the strategy $x_1^k$ and of the Lagrangian multiplier $\lambda_1^k$ of player $1$, (b) Trajectory of $\parallel \bld L \bld \lambda^k \parallel_2$, (d) The averaged constraints violation.}
%\label{fig:simulations}
%\end{figure}
\begin{figure}
\centering
\subfloat[][]
{\includegraphics[width=.35\textwidth]{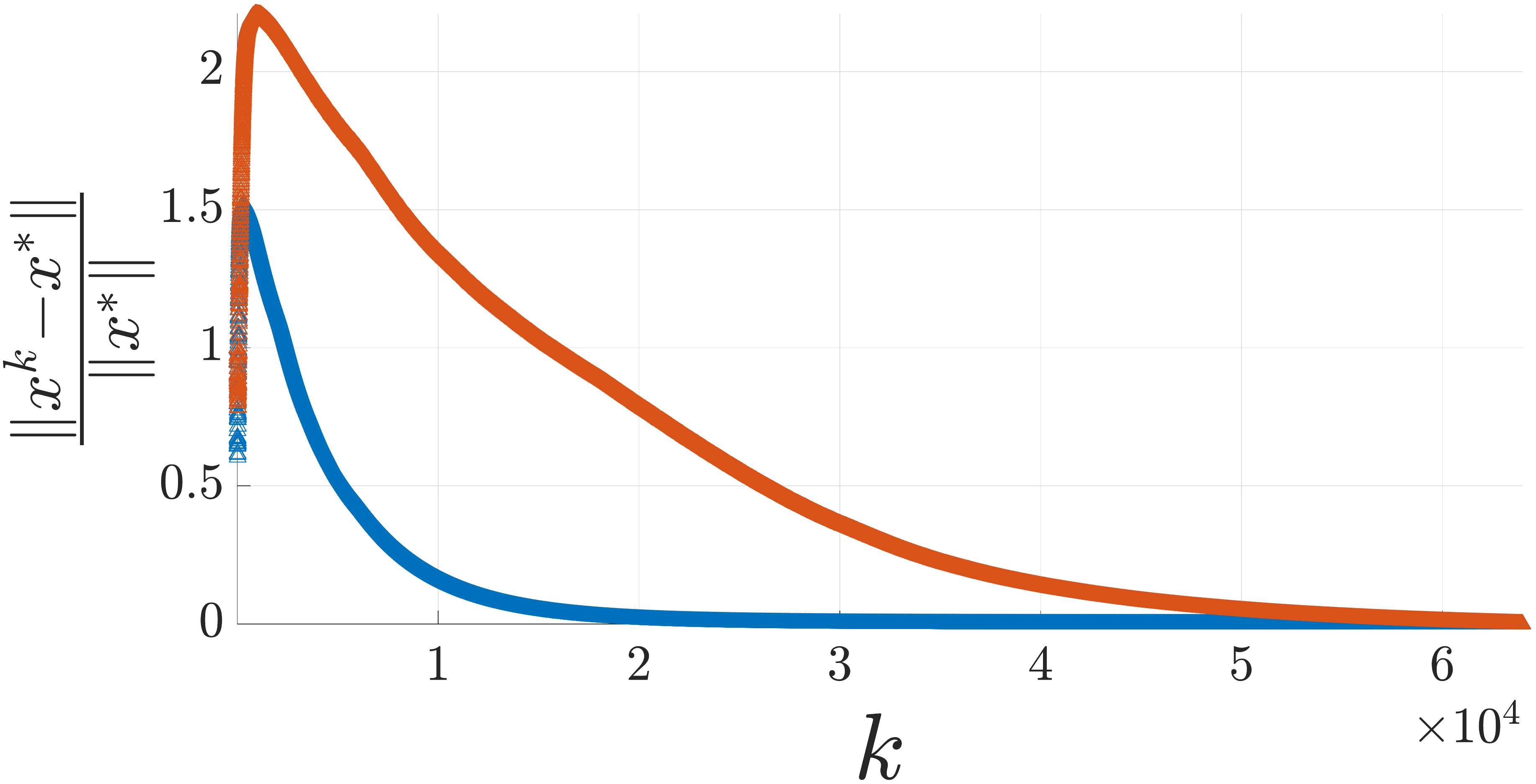}\label{fig:NormXVal}} \\
\subfloat[][]
{\includegraphics[width=.35\textwidth]{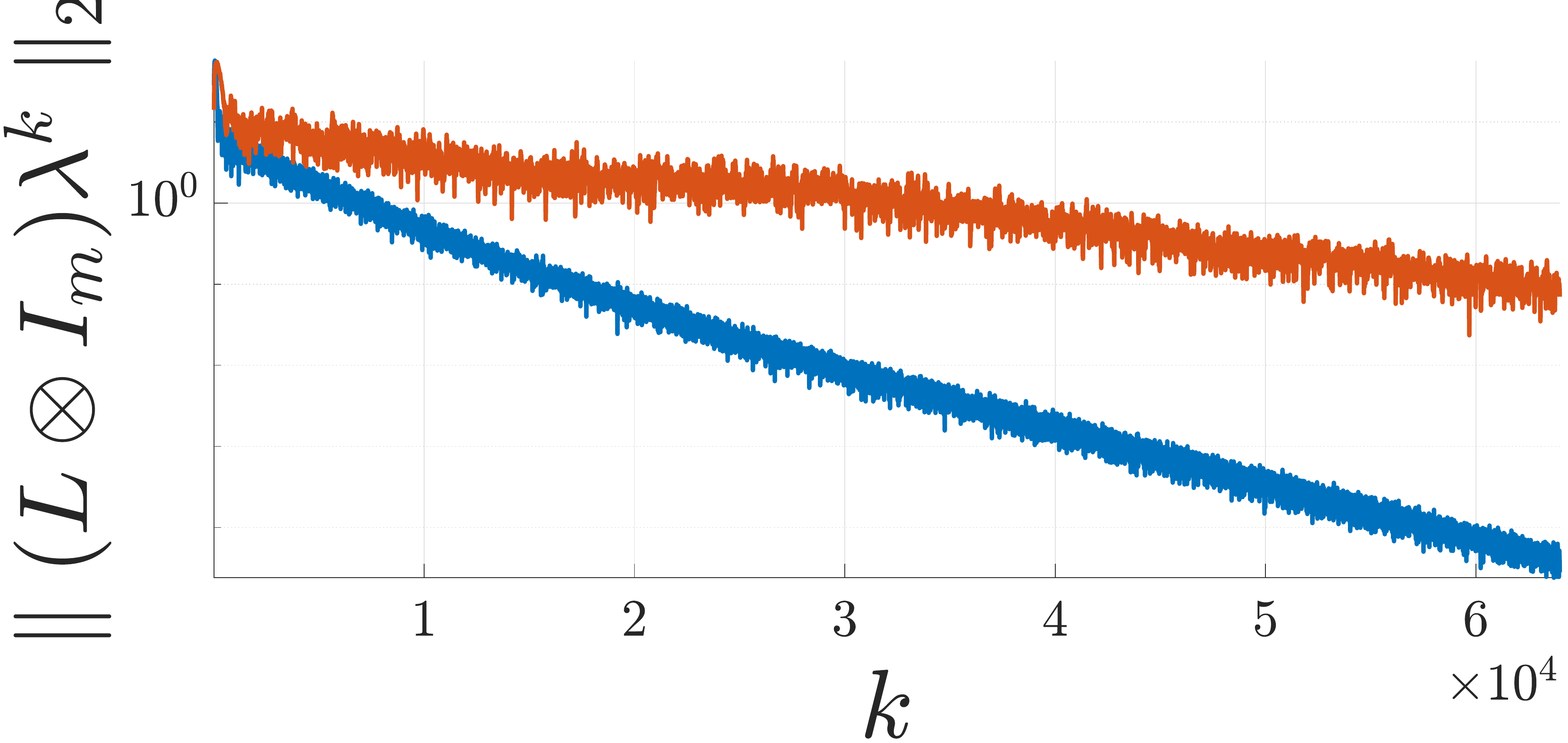}\label{fig:LambdaErr}} \\
\subfloat[][]
{\includegraphics[width=.35\textwidth]{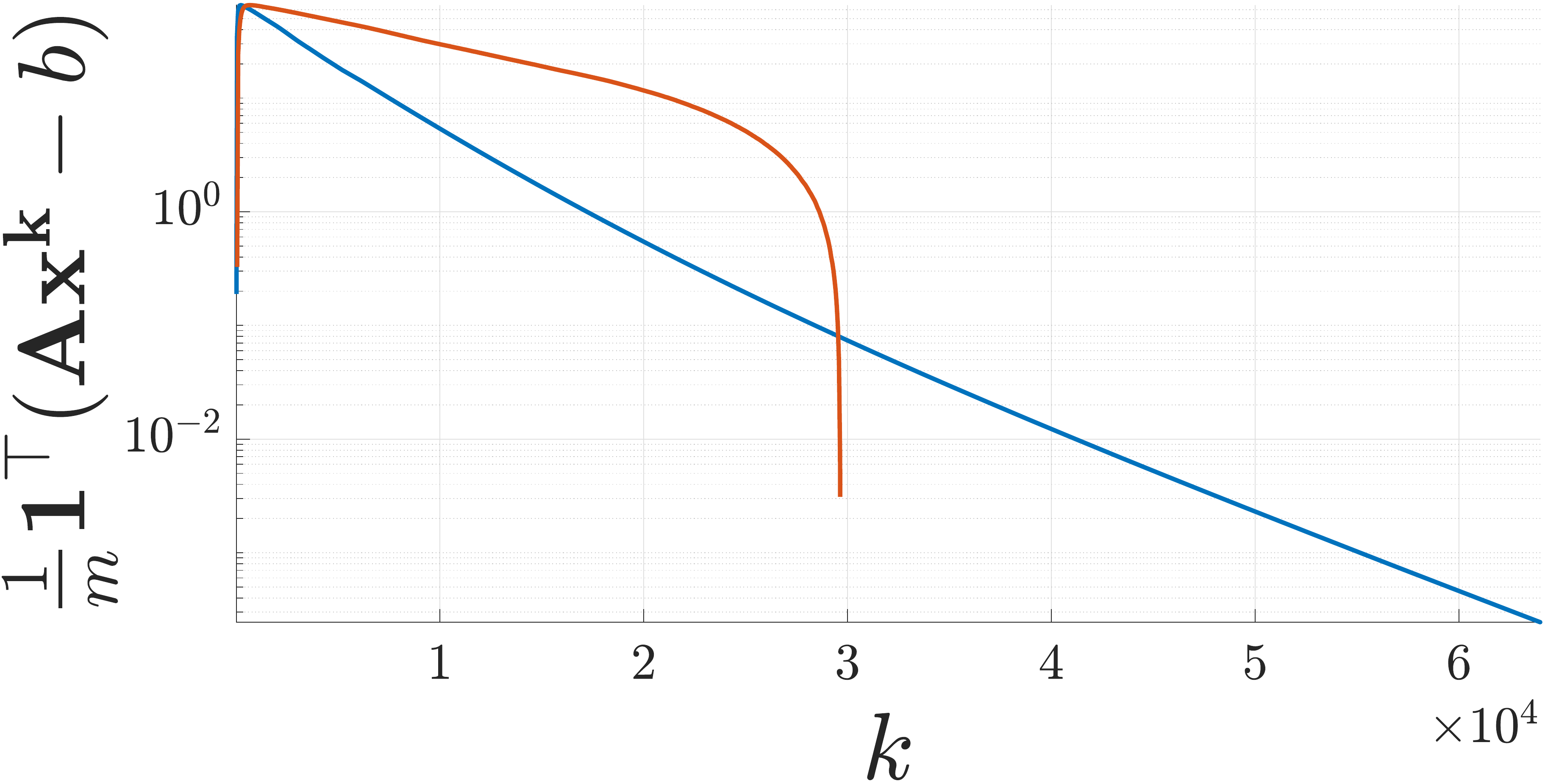}\label{fig:Constr}}\\
\caption{Communication in alphabetic order (blue) versus random communication (red): (a) Normalized distance from equilibrium,(b) Norm of the disagreement vector, (c) Averaged constraints violation (the negative values are omitted).}
\label{fig:simulations}
\end{figure}
We propose two setups, the case of communication over a ring graph with alphabetic order and the case of random communication (in Figure~\ref{fig:simulations}, respectively the blue and red trajectories). In the latter, we only ensure that every $20N$ iterations all the agents performed a similar number of updates. The edges of the graph are arbitrarily oriented. We assume that the agents  update with uniform probability, i.e., $P[\zeta^k=\Upsilon_i]=\frac{1}{N}$. The step sizes $\delta,\:\varepsilon,\:\tau$ in AD-GENO  are randomly chosen, the first from $[0.5,0.2]$ and the others from $[0.5,0.03]$, in order to ensure $\Phi\succ 0$ and $\eta = 0.35$. The maximum delay is assumed $\bar \varphi = 4$, therefore $\hat\varpi^k$ in \eqref{eq:Krasno_asynch_final} is $\hat\varpi^k = \col(\hat \varpi_1^k,\dots,\hat \varpi_N^k)$ where each $\hat \varpi_i^k$ is randomly chosen from $\{\hat \varpi_i^{k-\bar \varphi},\dots,\hat \varpi_i^k\}$.

The results of the simulations are shown in Figure~\ref{fig:simulations}. In particular, Figure~\ref{fig:NormXVal} presents the convergence of the collective strategy $\bld x^k$  to the v-GNE $\bld x^*$. Furthermore, Figure~\ref{fig:LambdaErr} highlights the convergence of the Lagrangian multipliers to consensus. We noticed that a simple update sequence, as the alphabetically ordered one, leads to a faster convergence than a  random one. In general, the more the agents' updates are well mixed the faster the algorithm converge.  

\section{Conclusion}
\label{sec:conclusion}

%
%A suitable variant of the forward-backward splitting algorithm can solve generalized Nash equilibrium problems via asynchronous, distributed information exchange, robustly to  communication delays.
%

%A change of variables based on the edge Laplacian matrix of the information-exchange graph allows one to preserve scalability of the solution algorithm in the number of nodes (as opposed to the number of edges).
%
%Full theoretical and numerical comparison between the proposed solution algorithm and that in [15] is left as future work.

This work propose a variant of the forward-backward splitting algorithm to solve generalized Nash equilibrium problems via asynchronous and distributed information exchange, that is robust to  communication delays.
A change of variables based on the node Laplacian matrix of the information-exchange graph allows one to preserve the scalability of the solution algorithm in the number of nodes (as opposed to the number of edges).
Full theoretical and numerical comparison between the proposed solution algorithm and that in \cite{Pavel:Yi:2018:Asynch} is left as future work.
Another interesting topic is the adaptation of the algorithm to the case of changing graph topology, in fact the independence from the edge variables makes this approach more suitable to address this problem.

%In this paper, we proposed an asynchronous distributed iterative algorithm for the computation of variational GNE of noncooperative games with affine coupling constraints. It does not require the use of a global synchronized clock since it is robust to delayed neighbors information. The design is based on a preconditioned forward backward splitting and adopts only node auxiliary variables to ensure the convergence to consensus of the Lagrangian multipliers, while preserving the scalability. The theoretical results provide a bound over the step sizes of the algorithm to ensure convergence. We numerically test our results solving a Cournot competition game in network markets.
%
%The simulations were run on a single computer, taking away the change of real comparison between computational time of the synchronous and asynchronous algorithm. A future work will be  an implementation of these solutions on a real distributed system to deeply understand the benefit of both approaches. Furthermore, another interesting research topic is a detailed comparison between the proposed algorithms and the one developed in \cite{Pavel:Yi:2018:Asynch} to study the upsides and downsides of the each algorithms.  
%
%
%%Moreover, the independence to the edges makes the algorithm more suitable for future adaptation to the case of changing graph topology.  

%%%%%%%%%%%%%%%%%%%%%%%%%%%%%%

\appendix 
\subsection*{Proof of Lemma~\ref{lem:max_mon_of_operators}}
It follows as in \cite[Proof of Th.~4.5]{pavel2017:distributed_primal-dual_alg}.
%\alert{Note:can be omitted and refer to Pavel}
%
%The operator $\ca A$ is the sum of two operators, the first is a real skew symmetric matrix, that is maximally monotone by \cite[Ex.~20.35]{bauschke2011convex}. The second instead is $0\times N_\Omega\times N_{\bR^{mN}_+}$, maximally monotone from \cite[Ex.~20.26]{bauschke2011convex}. Finally, invoking \cite[Cor.~25.5]{bauschke2011convex}, we conclude that $\ca A$ is maximally monotone.
%
%By Assumption~\ref{ass:subgrad_lipschitz_strong_mon}, we can use the definitions of strong monotonicity and Lischitz continuity to complete the proof. Given two points $\varpi_1,\varpi_2$ it always holds that $\langle \ca B(\varpi_1)-\ca B(\varpi_2),\varpi_1-\varpi_2\rangle=\langle  F(\bld x_1)- F(\bld x_2),\bld x_1-\bld x_2\rangle\geq \alpha \lVert \bld x_1-\bld x_2\rVert \geq \frac{\alpha}{\ell^2} \lVert F(\bld x_1)-F(\bld x_2)\rVert =\frac{\alpha}{\ell^2} \lVert \ca B(\bld x_1)-\ca B(\bld x_2)\rVert $, hence invoking \cite[Def.~4.10]{bauschke2011convex} $\ca B$ is $\frac{\alpha}{\ell^2} $-cocoercive.
\hfill\QED
\smallskip
\subsection*{Proof of Theorem~\ref{th:eq_are_vGNE_mod_map}}
Given the equilibrium point $\col (\bld x^*,\bld \lambda^*,\bld z^*)$,  \eqref{eq:row1_mod_sync} evaluated in the equilibrium reduces to $\bld 0 = \bld L\bld \lambda^*$, that implies $\bld \lambda^* = \lambda^*\otimes \bld 1$. 

Moreover manipulating \eqref{eq:row2_sync_mod} and evaluating it in the equilibrium, we obtain 
\begin{equation}
\label{eq:row2_proof_equilibrium}
\bld 0 \in N_{\bR^{mN}_{\geq 0}}(\bld \lambda^*)+\bar b-\Lambda \bld x^* + 2\bld \delta \bld L \bld \lambda^* +\bld z^* \,.
\end{equation}
Exploiting the property $\bld L \bld \lambda^* = \bld 0$ and multiplying both side of  \eqref{eq:row2_proof_equilibrium} by $(\bld 1^\top \otimes I)$ leads to
\begin{equation}
\label{eq:row2_proof_equilibrium_2}
\bld 0 \in  (\bld 1^\top \otimes I)(N_{\bR^{mN}_{\geq 0}}(\bld \lambda^*)+\bar b-\Lambda \bld x^* +\bld z^* )\,.
\end{equation}   
Using the fact that $\sum_{i\in\ca N}N_{\bR^m_{\geq 0}} (\lambda^*)=N_{\cap_{i\in\ca N} \bR^m_{\geq 0}} (\lambda^*)=N_{\bR^m_{\geq 0}} (\lambda^*)$ and the assumption $\bld 1^\top \bld z^* =0$,  \eqref{eq:row2_proof_equilibrium_2} becomes
\begin{equation}
\label{eq:row2_proof_equilibrium_3}
\bld 0 \in  N_{\bR^{m}_{\geq 0}}(  \lambda^*)+\bar b - A \bld x^*\,.
\end{equation}

Finally, \eqref{eq:row3_sync} evaluated in the equilibrium is 
\begin{equation}
\label{eq:row3_proof_equilibrium_2}
\bld 0 \in F(\bld x^*)+N_{\bld \Omega}(\bld x^*)+\Lambda^\top \bld \lambda^*\\
\,,
\end{equation}   
or equivalently
\begin{equation}
\label{eq:row3_proof_equilibrium_3}
\bld 0 \in  \nabla f_i(x_i^*,\bld x^*_{-i})+N_{\Omega_i}( x^*_i)+A_i^\top  \lambda^*\:,\quad \forall i\in\ca N\,.
\end{equation}
Inclusions \eqref{eq:row3_proof_equilibrium_3} and \eqref{eq:row2_proof_equilibrium_3} are the KKT conditions in \eqref{eq:KKT_VI}, hence from \cite[Th.~3.1]{facchinei:fischer:piccialli:07} we conclude that $\col (\bld x^*,\bld \lambda^*,\bld z^*)$ is a v-GNE of the game.
\hfill\QED
\smallskip
\subsection*{Proof of Theorem~\ref{th:convergence_sync}}
First, we note from \cite[Lem.~5.6]{pavel2017:distributed_primal-dual_alg} that  the two operators $\Phi^{-1}\ca B$ and $\Phi^{-1}\ca A$ are respectively $\frac{\alpha\vartheta}{\ell^2}$-cocoercive and  maximally monotone. Furthermore, this also implies that $(\Id-\Phi^{-1}\ca B)$ is $\frac{\ell^2}{2\alpha\vartheta}$-AVG and $\mathrm{J}_{\Phi^{-1}\ca A}=(\Id+\Phi^{-1}\ca A)$ is FNE. Applying \cite[Prop.~2.4]{combettes:yamada:15}, we conclude that $T$ is $\frac{2\alpha\vartheta}{4\alpha\vartheta-\ell^2}$-AVG. The Krasnosel'ski\u i iteration in \eqref{eq:Krasno_iter_synch} converges to $\varpi^*\in\fix(T)$ if $\eta\in(0,\frac{4\alpha\vartheta-\ell^2}{2\alpha\vartheta})$, \cite[Th.~5.14]{bauschke2011convex}. 

From the above argument, it holds that $\lim_{k\rightarrow +\infty}\bld \sigma^k = \bld \sigma^*$, hence $\lim_{k\rightarrow +\infty} \bld E^\top \bld \sigma^k = \bld E^\top \bld \sigma^*=:\bld z^*$, therefore also $\bld z$ converges. Therefore, we conclude that Algorithm~\ref{alg:synch_alg} converges to $\col(\bld x^*,\bld \lambda^*,\bld z^*)$. The choice of the initial value $\bld z_0=\bld 0$, implies that $\bld 1^\top \bld z^k =0$, $\forall k\in\bN_{\geq 0}$ since its values will in the range of the Laplacian matrix.
 Finally, applying Theorem~\ref{th:eq_are_vGNE_mod_map} we conclude that the equilibrium is the v-GNE of the original game.
\hfill\QED
\smallskip
\subsection*{Proof of Theorem~\ref{th:convergence_asynch_sigma}}
From the proof of Theorem~\ref{th:convergence_sync}, we know that $T$ is $\frac{2\alpha\vartheta}{4\alpha\vartheta-\ell^2}$-AVG, therefore is can be rewritten as $T=(1-\frac{2\alpha\vartheta}{4\alpha\vartheta-\ell^2})\Id+\frac{2\alpha\vartheta}{4\alpha\vartheta-\ell^2}  P$, where $P$ is nonexpansive, \cite[Prop.~4.35]{bauschke2011convex}. By substituting it into \eqref{eq:Krasno_asynch_final}, we obtain
\begin{equation}\label{eq:Krasno_asynch_rule_proof}
\varpi^{k+1}=\varpi^k+ \frac{2\eta\alpha\vartheta}{4\alpha\vartheta-\ell^2} \zeta^k( P-\Id)\hat\varpi^k\,.
\end{equation}
For \eqref{eq:Krasno_asynch_rule_proof}, we apply \cite[Lem.~13 and Lem.~14]{peng2016arock}, to conclude that $\{\varpi^k\}_{k\in\bN_{\geq 0}}$ is bounded and that it converges almost surely to $\varpi^*\in\fix(P)=\fix(T)$, for $\frac{2\eta\alpha\vartheta}{4\alpha\vartheta-\ell^2}\in(0,\frac{cNp_{\min}}{2\bar\varphi\sqrt{p_{\min}}+1}]$ with $p_{\min}:= \min\{p_i\}_{i\in\ca N}$. 
Since $\fix(T)=\zer(\ca A+\ca B)$, we conclude from Proposition~\ref{prop:zer_AB_are_vGNE} that $\{\bld x^k\}_{k\in\bN_{\geq 0}}$ converges to the v-GNE of \eqref{eq:game_formulation} almost surely.
\hfill\QED
\smallskip
\subsection*{Proof of Theorem~\ref{th:convergence_AD-GENO}}
The change of auxiliary variables in Algorithm~\ref{alg:AD-GENO} from $\bld \sigma$ to $\bld z$ leads to a different update rule for $\bld \lambda$, instead the one for $\bld x$ remains unchanged. Therefore, if we show that the new update of $\bld \lambda$ is equivalent to the one in Algorithm~\ref{alg:E-ADAGNES}, we can infer the convergence from Theorem~\ref{th:convergence_asynch_sigma}.

\smallskip
We prove by induction that, given an agent $i$, the update of $\lambda_i$ at time $k$ in Algorithm~\ref{alg:E-ADAGNES}~and~\ref{alg:AD-GENO} are equivalent. Note that the two update rules are equivalent if it holds that 
\begin{equation}\label{eq:condition_z_sigma}
\tilde z_{i}^{k}=([E^\top]_{i}\otimes I_m)\hat{ \bld \sigma}^{k}\:,
\end{equation} 
for every $k>0$ and $i\in\ca N$.

\underline{\textit{Base case:}} Iteration $k$ is the first in which agent $i$ updates its variables. 
If $k=0$, then $\mu_i =\bld 0_{m}$ in AD-GENO and $\hat{\bld \sigma}^0 = \bld 0_{mM}$ in E-ADAGNES, hence \eqref{eq:condition_z_sigma} is trivially verified.

If instead $k>0$, it holds that $z_{i}^{k-\varphi^k_i}=z_{i}^0=\bld 0_m$, while $\hat{\bld \sigma}^k \not = \bld 0_{mM}$, since the neighbours of $i$ can update more than once before the first update of $i$. We define for each $j\in\ca N_i$ the set $\ca S_{j}^k$, a $t\in\bN$ where $t<k$ belongs to $\ca S_{j}^k$ if at the iteration $t$ of the algorithm agent $j$ completes an update. %For simplicity, we also assume that its elements are in a decreasing order. 
We define the maximum time in $\ca S_{j}^k$ as $m_j^k:=\max\{t\:|\:t\in\ca S_{j}^k\}$ and $\check {\ca S}_{j}^k:={\ca S}_{j}^k\setminus m_j^k$.
From this definitions, we obtain that   
\begin{equation}\label{eq:E_top_sigma_hat}
([E^\top]_{i}\otimes I_m)\hat{ \bld \sigma}^{k} = \textstyle{ \sum_{l\in\ca E^{\mathrm{out}}_i} } \sigma_{l}^0 - \textstyle{ \sum_{d\in\ca E^{\mathrm{in}}_i} } \sigma_{d}^{m_j^k}\:,
\end{equation}      
where $j$ is the element of $e_d$ different from $i$. Furthermore, from the update rule of $\sigma_d$ in Algorithm~\ref{alg:E-ADAGNES}, we derive
\begin{equation}\label{eq:definition_sigma_d}
\begin{split}
 \sigma_{d}^{m_j^k} &= -\sigma_{l}^{\max\{t\:|\:t\in\check{\ca S}_{j}^k\}} - \eta\delta (\lambda_{j}^{\max\{t\:|\:t\in\check{\ca S}_{j}^k\}} - \lambda_{i}^0)\\
 & = \eta\delta\: \textstyle{ \sum_{h\in\check{\ca S}_{j}^k } }\left(  \lambda_{i}^0 - \lambda_{j}^h \right) 
\end{split}
\end{equation}
Substituting  \eqref{eq:definition_sigma_d} into \eqref{eq:E_top_sigma_hat} leads to
\begin{equation}\label{eq:E_top_sigma_hat2}
\begin{split}
([E^\top]_{i}\otimes I_m)\hat{ \bld \sigma}^{k} &= \eta \frac{\delta}{2}\left( \left[\textstyle{\sum_{j\in\ca N_i\setminus \{i \}} }|\check{S}_{j}^k| \right]\lambda_{i}^0 \right. \\ & \left.\qquad \qquad -  \textstyle{\sum_{j\in\ca N_i\setminus \{i \}} \sum_{h\in\check{\ca S}_{j}^k}} \lambda_{j}^h \right)\:.\\
\end{split}
\end{equation}
From the definition given in Algorithm~\ref{alg:AD-GENO} of $\mu_i$, we attain that $([E^\top]_{i}\otimes I_m)\hat{ \bld \sigma}^{k}=\eta \delta\mu_i = \tilde z_i^k $, therefore \eqref{eq:condition_z_sigma} hold.

\smallskip
\underline{\textit{Induction step:}} Suppose that \eqref{eq:condition_z_sigma} holds for some $\bar k >0 $ that corresponds the latest iteration in which agent $i$ performed the update, i.e. $z_{i}^{\bar k}\not = \bld 0$. 

Consider the next iteration $k$ in which agent $i$ updates, $k>\bar k$. Here, $\ca S_{j}^k$ is defined as above, but for time indexes $(\bar k,k]$. Following a similar reasoning to the previous case, we obtain
\begin{equation}\label{eq:E_top_sigma_hat_induction}
\begin{split}
([E^\top]_{i}&\otimes I_m)\hat{ \bld \sigma}^{k} =([E^\top]_{i}\otimes I_m)\hat{ \bld \sigma}^{\bar k} + \eta\delta\sum_{l\in\ca E_i^{\mathrm{out}}}([E]_{l}\otimes I_m)\hat{ \bld \lambda}^{\bar k}   \\ 
& +\eta \delta \left( \left[ \textstyle{ \sum_{j\in\ca N_i\setminus \{i \}} }|\check{S}_{j}^k| \right]\lambda_{i}^0 -  \textstyle{\sum_{j\in\ca N_i\setminus \{i \}} \sum_{h\in\check{\ca S}_{j}^k} }\lambda_{j}^h \right)\\
\end{split}
\end{equation}
where we used the fact that $l\in\ca E_i^{\mathrm{out}}$ is updated at the same time of $i$. Furthermore, from the induction assumption,
\begin{equation}
\begin{split}
([E^\top]_{i}\otimes I_m)\hat{ \bld \sigma}^{k} &=   z^{\bar k}_i +\eta \delta \left( \left[\textstyle{\sum_{j\in\ca N_i\setminus \{i \}} } |\check{S}_{j}^k| \right]\lambda_{i}^0 \right. \\
&\qquad  \left. -  \textstyle{\sum_{j\in\ca N_i\setminus \{i \}} \sum_{h\in\check{\ca S}_{j}^k} } \lambda_{j}^h \right)\\
& = z^{\bar k}_i +\eta \delta \mu_i= \tilde z^{ k}_i\:,
\end{split}
\end{equation}
where the last step holds because in the reading phase of Algorithm~\ref{alg:AD-GENO}, we reset to zero the values of $\mu_i$, every time $i$ starts an update. 
Therefore, \eqref{eq:condition_z_sigma} holds for $k$. 

This concludes the proof by induction, showing that the update of $\lambda_{i}$ defined in Algorithm~\ref{alg:AD-GENO} is equivalent to the one in Algorithm~\ref{alg:E-ADAGNES}, for any $k$. Finally, since the pair $(\bld x,\bld \lambda)$ has an update rule equivalent to the one in E-ADAGNES, the convergence of $\{\bld x\}_{k\in\bN_{\geq 0}}$ to the v-GNE of the game \eqref{eq:game_formulation} follows by Theorem~\ref{th:convergence_asynch_sigma}.     
\hfill\QED
%%%%%%%%%%%%%%%%%%% biblio %%%%%%%%%%%%%%
\bibliographystyle{IEEEtran}
\bibliography{libraryCC,librarySG}

% Generated by IEEEtran.bst, version: 1.14 (2015/08/26)
\begin{thebibliography}{10}
\providecommand{\url}[1]{#1}
\csname url@samestyle\endcsname
\providecommand{\newblock}{\relax}
\providecommand{\bibinfo}[2]{#2}
\providecommand{\BIBentrySTDinterwordspacing}{\spaceskip=0pt\relax}
\providecommand{\BIBentryALTinterwordstretchfactor}{4}
\providecommand{\BIBentryALTinterwordspacing}{\spaceskip=\fontdimen2\font plus
\BIBentryALTinterwordstretchfactor\fontdimen3\font minus
  \fontdimen4\font\relax}
\providecommand{\BIBforeignlanguage}[2]{{%
\expandafter\ifx\csname l@#1\endcsname\relax
\typeout{** WARNING: IEEEtran.bst: No hyphenation pattern has been}%
\typeout{** loaded for the language `#1'. Using the pattern for}%
\typeout{** the default language instead.}%
\else
\language=\csname l@#1\endcsname
\fi
#2}}
\providecommand{\BIBdecl}{\relax}
\BIBdecl

\bibitem{dorfer:simpson-porco:bullo:16}
F.~D\"{o}rfler, J.~Simpson-Porco, and F.~Bullo, ``Breaking the hierarchy:
  Distributed control and economic optimality in microgrids,'' \emph{IEEE
  Trans. on Control of Network Systems}, vol.~3, no.~3, pp. 241--253, 2016.

\bibitem{parise:colombino:grammatico:lygeros:14}
F.~Parise, M.~Colombino, S.~Grammatico, and J.~Lygeros, ``Mean field
  constrained charging policy for large populations of plug-in electric
  vehicles,'' in \emph{Proc. of the IEEE Conference on Decision and Control},
  Los Angeles, California, USA, 2014, pp. 5101--5106.

\bibitem{grammatico:18tcns}
S.~Grammatico, ``Proximal dynamics in multi-agent network games,'' \emph{IEEE
  Trans. on Control of Network Systems
  \textup{\url{doi.org/10.1109/TCNS.2017.2754358}}}, 2018.

\bibitem{martinez:bullo:cortes:frazzoli:07}
S.~Mart\'{i}nez, F.~Bullo, J.~Cort\'{e}s, and E.~Frazzoli, ``On synchronous
  robotic networks -- {Part} i: Models, tasks, and complexity,'' \emph{IEEE
  Trans. on Automatic Control}, vol.~52, pp. 2199--2213, 2007.

\bibitem{pavel2017:distributed_primal-dual_alg}
P.~Yi and L.~Pavel, ``A distributed primal-dual algorithm for computation of
  generalized nash equilibria via operator splitting methods,'' in \emph{2017
  IEEE 56th Annual Conference on Decision and Control (CDC)}, Dec 2017, pp.
  3841--3846.

\bibitem{facchinei:fischer:piccialli:07}
F.~Facchinei, A.~Fischer, and V.~Piccialli, ``On generalized {Nash} games and
  variational inequalities,'' \emph{Operations Research Letters}, vol.~35, pp.
  159--164, 2007.

\bibitem{BERTSEKAS:1991:Survey_Asynch}
D.~P. Bertsekas and J.~N. Tsitsiklis, ``Some aspects of parallel and
  distributed iterative algorithms—a survey,'' \emph{Automatica}, vol.~27,
  no.~1, pp. 3 -- 21, 1991.

\bibitem{bertsekas:1989:parallel_optimization}
------, \emph{Parallel and distributed computation: numerical methods}.\hskip
  1em plus 0.5em minus 0.4em\relax Prentice hall Englewood Cliffs, NJ, 1989,
  vol.~23.

\bibitem{recht:2011:hogwild}
B.~Recht, C.~Re, S.~Wright, and F.~Niu, ``Hogwild: A lock-free approach to
  parallelizing stochastic gradient descent,'' in \emph{Advances in neural
  information processing systems}, 2011, pp. 693--701.

\bibitem{Combettes:2015:stoch_quadi_fejer}
P.~Combettes and J.~Pesquet, ``Stochastic quasi-fejér block-coordinate fixed
  point iterations with random sweeping,'' \emph{SIAM Journal on Optimization},
  vol.~25, no.~2, pp. 1221--1248, 2015.

\bibitem{liu:2015:asynchronous_parallel_stoch_coord_desc}
J.~Liu, S.~J. Wright, C.~R{\'e}, V.~Bittorf, and S.~Sridhar, ``An asynchronous
  parallel stochastic coordinate descent algorithm,'' \emph{The Journal of
  Machine Learning Research}, vol.~16, no.~1, pp. 285--322, 2015.

\bibitem{nedic:2011:asynchronous_broadcast-based_convex_opt}
A.~Nedic, ``Asynchronous broadcast-based convex optimization over a network,''
  \emph{IEEE Transactions on Automatic Control}, vol.~56, no.~6, pp.
  1337--1351, 2011.

\bibitem{peng2016arock}
Z.~Peng, Y.~Xu, M.~Yan, and W.~Yin, ``Arock: an algorithmic framework for
  asynchronous parallel coordinate updates,'' \emph{SIAM Journal on Scientific
  Computing}, vol.~38, no.~5, pp. A2851--A2879, 2016.

\bibitem{Pavel:Yi:2018:Asynch}
P.~{Yi} and L.~{Pavel}, ``{Asynchronous distributed algorithms for seeking
  generalized Nash equilibria under full and partial decision information},''
  \emph{ArXiv e-prints}, Jan. 2018.

\bibitem{Paccagnan_Gentile2016:Distributed_computation_GNE}
D.~Paccagnan, B.~Gentile, F.~Parise, M.~Kamgarpour, and J.~Lygeros,
  ``Distributed computation of generalized nash equilibria in quadratic
  aggregative games with affine coupling constraints,'' pp. 6123--6128, Dec
  2016.

\bibitem{kulkarni:shanbhag:12}
A.~A. Kulkarni and U.~Shanbhag, ``On the variational equilibrium as a
  refinement of the generalized {Nash} equilibrium,'' \emph{Automatica},
  vol.~48, pp. 45Ð--55, 2012.

\bibitem{Facchinei:2011:KKT_and_GNE}
A.~Dreves, F.~Facchinei, C.~Kanzow, and S.~Sagratella, ``On the solution of the
  kkt conditions of generalized nash equilibrium problems,'' \emph{SIAM Journal
  on Optimization}, vol.~21, no.~3, pp. 1082--1108, 2011.

\bibitem{BELGIOIOSO:2017:convexity_and_monotonicity_aggr_games}
G.~Belgioioso and S.~Grammatico, ``On convexity and monotonicity in generalized
  aggregative games,'' \emph{IFAC-PapersOnLine}, vol.~50, no.~1, pp. 14\,338 --
  14\,343, 2017, 20th IFAC World Congress.

\bibitem{facchinei:pang}
F.~Facchinei and J.~Pang, \emph{Finite-dimensional variational inequalities and
  complementarity problems}.\hskip 1em plus 0.5em minus 0.4em\relax Springer
  Verlag, 2003.

\bibitem{Godsil:algebraic_graph_theory}
C.~{Godsil} and G.~{Royle}, \emph{Algebraic Graph Theory}, ser. Graduate Texts
  in Mathematics.\hskip 1em plus 0.5em minus 0.4em\relax Springer Science \&
  Business Media, 2013, vol. 207.

\bibitem{eckstein1989:splitting}
J.~Eckstein, ``Splitting methods for monotone operators with applications to
  parallel optimization,'' Ph.D. dissertation, Massachusetts Institute of
  Technology, 1989.

\bibitem{bauschke2011convex}
H.~H. Bauschke, P.~L. Combettes \emph{et~al.}, \emph{Convex analysis and
  monotone operator theory in Hilbert spaces}.\hskip 1em plus 0.5em minus
  0.4em\relax Springer, 2011, vol. 408.

\bibitem{yu:2017distributed}
C.-K. Yu, M.~van~der Schaar, and A.~H. Sayed, ``Distributed learning for
  stochastic generalized nash equilibrium problems,'' \emph{IEEE Transactions
  on Signal Processing}, vol.~65, no.~15, pp. 3893--3908, 2017.

\bibitem{combettes:yamada:15}
P.~L. Combettes and I.~Yamada, ``Compositions and convex combinations of
  averaged nonexpansive operators,'' \emph{Journal of Mathematical Analysis and
  Applications}, pp. 55--70, 2015.

\end{thebibliography}

\end{document}